\documentclass[12pt,letterpaper]{article}
\usepackage{booktabs,multirow}  
\usepackage{graphicx,lscape}
\usepackage{amsmath,amssymb}
\usepackage{mathtools}
\usepackage{bm}
\usepackage{threeparttable}
\usepackage{url}
\usepackage{graphicx}
\usepackage{soul}

\usepackage{color}

\usepackage{ulem} 

\usepackage{comment}

\newcommand{\blue}{\color{blue}}

\def\mybox#1{{\blue \vskip1mm \begin{center}
 \hspace{.0\textwidth}\vbox{\hrule\hbox{\vrule\kern6pt
  \parbox{.8\textwidth}{\kern6pt#1\vskip6pt}\kern6pt\vrule}\hrule}
  \end{center} \vskip-5mm}}

\begin{document}
\baselineskip=26pt

\begin{center}
\Large \bf Frequentist analysis of basket trials with one-sample Mantel-Haenszel procedures
\end{center}

\vspace{3mm}

\begin{center}
Satoshi Hattori \\
{\it Department of Biomedical Statistics, Graduate School of Medicine and \\
Integrated Frontier Research for Medical Science Division, Institute for Open and Transdisciplinary ResearchInitiatives (OTRI), Osaka University \\
Yamadaoka 2-2, Suita City, Osaka 565-0871, Japan \\
E-mail:hattoris@biostat.med.osaka-u.ac.jp 
}\\
\vspace{6mm}
Satoshi Morita \\
{\it 
Department of Biomedical Statistics and Bioinformatics, \\ 
Kyoto University Graduate School of Medicine, \\
54 Kawahara-cho, Shogoin, Sakyo-ku, Kyoto 606-8507, Japan. \\
E-mail: smorita@kuhp.kyoto-u.ac.jp
}
\end{center}

\begin{center}
Running title: Frequentist analysis of basket trials
\end{center}

\begin{center}
Version: 16Feb2023
\end{center}

\abstract{
Recent substantial advances of molecular targeted oncology drug development is requiring new paradigms for early-phase clinical trial methodologies to enable us to evaluate efficacy of several subtypes simultaneously and efficiently. The concept of the basket trial is getting of much attention to realize this requirement borrowing information across subtypes, which are called baskets. Bayesian approach is a natural approach to this end and indeed the majority of the existing proposals relies on it. On the other hand, it required complicated modeling and may not necessarily control the type 1 error probabilities at the nominal level. 
In this paper, we develop a purely frequentist approach for basket trials based on one-sample Mantel-Haenszel procedure relying on a very simple idea for borrowing information under the common treatment effect assumption over baskets. We show that the proposed estimator is consistent under two limiting models of the large strata and sparse data limiting models (dually consistent) and propose dually consistent variance estimators. The proposed Mantel-Haenszel estimators are interpretable even if the common treatment assumptions are violated. Then, we can design basket trials in a confirmatory matter. We also propose an information criterion approach to identify effective subclass of baskets. 
}

Key words: A key word; Dual consistency; Generalized information criterion; Mantel-Haenszel estimator; Oncology

\section{Introduction}
\label{s:intro}

Traditionally, clinical development of oncology drugs were made separately by cancer types such as breast cancer, lung cancer and so on, or more specifically like {\it Her2}-positive breast cancer and non-small cell lung cancer. In this traditional paradim, the phase 2 studies were usually conducted as a single-arm study with the objective response as the primary outcome focusing on a specific cancer type to evaluate whether the drug would be promising in the successive confirmatory large-scale phase 3 studies. Usually, it is designed to judge whether the objective response rate is superior to historical controls with a sufficient power. 

Recent substantial advances in molecular-targeted oncology drugs have suggested that some molecular-targeted drugs may be expected to be effective across several cancer types sharing similar signal pathways. It has motivated us to make clinical developments of such drugs quickly and efficiently by evaluating efficacy for several cancer types or subtypes simultaneously in a single study. The basket trial design is an attempt to realize this requirement, in which cancer types or subtypes of a single cancer type are called baskets. 
If sufficient number of patients could be enrolled in all the baskets, the parallel execution of the standard single-arm phase 2 studies may be a reasonable choice. 

Recently, the basket-trial design is getting of great interst; several papers in clinical journals addressed the issue and introduced this new design (LeBlanc, Rankin and Crowley 2009; Renfro and Sargent 2016; Cunanan {\it et al.} 2017a; Kaizer {\it et al.} 2019 among others). 

If some baskets share the common or similar biological mechanisms, one may expect to borrow information across baskets to make an efficient inference. At the design-stage, the investigators try to select baskets of homogeneous efficacy.  However, it may suffer from severe uncertainty and then the resulting efficacies may be highly variable over baskets. Then, it poses a very challenging question how we should borrow information across baskets. 

To address this issue, many Bayesian methods have been proposed to use Bayesian hierarchical models ($BHM)$, e.g., 
Thall {\it et al.} (2003) and Berry {\it et al.} (2013).
The key assumption of $BHM$ to borrow information across baskets is the exchangeablility across baskets. 
Neuenschwander {\it et al.} (2016) and Chu and Yuan (2018) discussed weakening the exchangeability assumption.
Although the shrinkage estimation under the $BHM$ could improve the reliability of basket-wise inference, the inflation of the type 1 error probability was discussed as one of the drawbacks of the Bayesian approaches to basket trials (Freidlin and Korn 2013; Zhou and Ji 2020; Jin {\it et al.} 2020). 
There are many literatures on the Bayesian approach such as Chen and Lee (2019) and Jin {\it et al.} (2020) who aimed to refine the $BHM$-based methods, Simon, {\it et al.} (2016) and Psioda {\it et al.} (2021) using Bayesian model-averaging, and Fujikawa {\it et al.} (2020) borrowing information across baskets based on similarity among posterior distributions. 
A comprehensive review of the methodology for the basket trial is given by Pohl, Krisam and Kieser (2021).

Comparing to huge accumulation of literatures on Bayesian basket trial methodologies, very limited number of frequentist approaches have been proposed. LeBlanc, Rankin and Crowley (2009) discussed a method to improve the power of the binomial test against a null response rate of clinically meaningful minimum response rate. In the first stage, futility stopping for ineffectiveness was considered with basket-specific binomial tests, and then combine all the baskets not stopped with the first-stage futility testing. This method was reported to have higher powers than the test simply combining all the baskets. Cunanan {\it et al.} (2017) made some refinements of the method by LeBlanc, Rankin and Crowley (2009); a step to judge homogeneity among baskets was introduced controlling the family wise type 1 error probability. LeBlanc, Rankin and Crowley (2009) and Cunanan {\it et al.} (2017) considered the null hypothesis of the common response rate over the baskets and took the approach to combine baskets of the seemingly similar overall response rate. This approach was referred as the "pooling all or nothing" approach in the recent review paper by Pohl, Krisam and Kieser (2021) and was the only frequentist approach referred therein. Krajewska and Rauch (2021) considered a frequentist extension of the "pooling all or nothing" approach to allow several subclasses of baskets to pool incorporating clustering techniques of baskets.

On the other hand, clinically meaningful minimum response rates can be heterogeneous among baskets. London and Chang (2005) allowed basket-specific null hypotheses. Instead of selecting baskets to pool, London and Chang (2005) proposed the unconditional and conditional tests for a global null hypothesis thoughout all the baskets with a single test statistic defined by the sum of the number of responders in each basket.  Chang, Shuster and Hou (2012) proposed a weighted version of the test statistic by London and Chang (2005) from the viewpoint of the most powerful test. Both London and Chang (2005) and Chang, Shuster and Hou (2012) considered two-stage procedures to allow early stopping. Since these methods were for a single global null hypothesis without any adaptive pooling of baskets, no alpha-level adjustment was not needed for the single-stage analysis and the type 1 error probabilities could be easily controlled for the two-stage methods. Despite of the advantage of the proper control of the type 1 error probabilities, the methods by London and Cheng (2005) and  Chang, Shuster and Hou (2012) did not address estimation of global or basket-specific estimation. 

In this paper, we propose a purely frequentist approach for basket trials, covering a series of inference; testing, estimation and identification of effective baskets. In Section 2, we give an overview of the proposed procedure and the insight behind it. In Section 3, we introduce the setting and notations, as well as a brief review of the existing frequentist approaches. For testing, we employ the conditional test by London and Cheng (2005) and for estimation, we introduce a new method, called the one-sample Mantel-Haenszel estimator in Section 4. For effective basket identification, we take a model-selection approach vis information criterion. In Section 5, we introduce the generalized information criterion for the one-sample Mantel-Haenszel estimator. Applications to real datasets are presented in Section 6 and in Section 7, we report results of the simulation studies to examine performance of the proposed procedure. We conclude the paper by making some remarks in Discussion section. All the theoretical details are placed in Appendix. 

\section{Overview of the proposal}
As reviewed, the Bayesian approaches have the advantage to address basket-wise inference borrowing information across baskets, but they have troubles to control the type 1 error probabilities. Furthermore, the Bayesian methods are computationally demanding and strongly rely on complicated model specifications. On the other hand, the frequentist approaches can control the type 1 error probabilities, but cannot necessarily address basket-specific inference well. Basically, existing frequentist approaches can address only testing hypothesis and do not address estimation in particular for baskets with very limited number of patients. In our perspective, the following three natures should be possessed by a basket trial design; \\
(I) It should be evaluated whether the test drug is effective at least for some baskets.  \\
(II) Suitable summary measures for overall efficacy are estimable, which can be interpreted without any strong assumptions. \\
(III) It can be addressed which baskets the test drug is effective for or not. \\
Although the phase 2 studies of oncology drugs have strong explanatory aspects, we believe that it is more appealing to design phase 2 studies with strong control of the type 1 error probabilities for a pre-specified null hypothesis with simple statistical methods as routinely done in the traditional single-arm phase 2 studies (Simon 1989).  
 For the item (I), the test by London and Chang (2005) provides a solution. We use their conditional exact test. To handle basket trials of  possibly very limited number of patients, borrowing information is a crucial step. Instead of taking the Bayesian shrinkage approach, we take a much simpler approach to pool information assuming the common effect-size parameters. To be specific, we take the Mantel-Haenszel (MH) approach, which is widely used in epidemiology to combine the common odds ratio (Breslow 1981; Robins, Breslow and Greenland 1986; Sato 1990) for stratified $2 \times 2$ tables. We propose one-sample version of the MH estimators for the risk difference (RD) and the risk ratio (RR) (Greenland and Robins 1985; Sato 1989) to the basket-specific null response rates. Same as the original Mantel-Haenszel estimators, the proposed one-sample MH estimator is shown to be dually consistent in the sense that it is consistent when the number of patients in each basket goes to infinity (large strata limiting model; Asymptotic 1) or the number of baskets goes to infinity (sparse data limiting model; Asymptotic 2). The construction of the one-sample MH estimator for the RD and that for the RR are on the common RD and RR assumptions, respectively. We show that even if the common RD assumption is violated, the MH risk difference estimator converges to the average of the basket-specific RD weighted by sample size. We also show that by putting a weight, the MH risk ratio estimator converges to the average of basket-specific RRs weighted by sample size even if the common RR assumption is violated. Thus, MH estimators are always interpretable. This is a preferable property for the primary statistical analysis method to be specified in the study protocol. 
Furthermore, we propose dually-consistent variance estimators for the MH estimators. These dually consistent asymptotic properties are desirable for basket trials since some baskets may have very limited number of patients. All the estimators and variance estimators have a simple closed-form expression, which do not require any complicated computation. Then, we can address the items (I) and (II) in a confirmatory matter. In words, with the prespecified methods of a simple closed-form expression, one can address whether the test drug is effective for at least one basket and the average treatment effect over baskets specified at the design-stage can be quantified with the one-sample MH estimators. We propose to address the item (III) in an explanatory matter. One may use any existing Bayesian methods to this end. We propose a simple frequentist approach. We developed the Generalized Information Criterion (GIC; Konishi and Kitagawa 1996) for this purpose. The information criterion is regarded a bias-adjusted Kullback-Leibler divergence. We show that the derived GIC is justified both under Asymptotic 1 and Asymptotic 2 or in the dual sense. We recommend a method to identify effective baskets by clustering baskets with the GIC.

\section{Setup, notation and existing frequentist tests}
Suppose we are interested in a single-arm oncology basket trial of $K$ baskets with a binary tumor response as the primary endpoint. For the $k$th basket ($k=1,2,...,K$), let $n_k$ and $Y_k$ be the number of patients enrolled and that responded to the treatment in the $k$th basket. Set $n=\sum_{k=1}^K n_k$. We assume that conditional on $n_k$, $Y_k$ follows the binomial distribution $Bin (n_k, \pi_k)$, where $\pi_k$ is the true response rate of the $k$th basket. Consider the following hypothesis testing problems; the null hypothesis is
\begin{eqnarray}
H_0: \pi_k = \pi_{k,0}
\label{null} 
\end{eqnarray} 
for any $k$, where $\pi_{k,0}$ is a basket-specific clinically meaningful minimum overall response rate, which should be set accounting for historical data and clinical perspective. A simple frequentist approach to this hypothesis testing is to apply the standard one-sample binomial test separately for each basket possibly with multiplicity adjustment. However, this approach is not appealing if some baskets have very limited number of patients. We consider the situation, in which the number of patient $n_k$ are not necessarily large in some baskets.
London and Chang (2005) discussed several tests for this hypothesis testing. We focus on their conditional test. 
Let $T=\sum_{k=1}^K Y_{k}$, or more generally $T_w=\sum_{k=1}^K w_k Y_{k}$ (Chang et al. 2012), where $w_k$ is a known weight. For RD, the constant weight $w_k=1$ is considered. For RR, $w_k=\pi_{k,0}^{-1}$ is suggested. The test statistic  with this weight is denoted by $T_iw=\sum_{k=1}^K \pi_{k,0}^{-1} Y_{k}$. See Section 4.3 on the choice of the weight. Under the null hypothesis, $T_w$ is a weighted sum of independent binomial random variables following $Bin (n_k, \pi_{k,0}); k=1,2,...,K$. The null distribution of $T_w$ can be evaluated easily by analytic calculation or generating sets of random numbers $\tilde{Y}_k \sim Bin (n_k, \pi_{k,0})$ and then calculating the empirical distribution of $\tilde{T}=\sum_{k=1}^K \tilde{Y}_k$.

\section{One-sample Mantel-Haenszel procedure}
\subsection{Mantel-Haenszel risk difference}
Let the RD of the $k$th basket denoted by $\Delta_k^{RD}=\pi_k-\pi_{k,0}$. 
In this subsection, we assume the RD is common across baskets and the common RD is denoted by $\Delta^{RD}$. That is, for $k=1,2,...,K$, 
\begin{eqnarray}
\Delta^{RD}=\Delta_k^{RD}=\pi_k-\pi_{k,0}. 
\label{rd1}
\end{eqnarray}
Denote the empirical overall response rate of the $k$th basket by $\hat{\pi}_{k}=Y_k/n_k$. Peplacing $\pi_k$ in $(\ref{rd1})$ with $\hat{\pi}_k$, the relationship
$n_k \Delta^{RD} \simeq Y_k-n_k \pi_{k,0}$ is suggested, where we use the symbol $\simeq$ to represent approximation in a rather informal matter to explain our motivation for the proposed method. Formal asymptotic justification is made. By summing them up,
\begin{eqnarray*}
\sum_{k=1}^{K} n_k \times \Delta^{RD} \simeq \sum_{k=1}^{K} (Y_k-n_k \pi_{k,0})
\label{rd2}
\end{eqnarray*}
is suggested. It motivates the estimator for the common RD,
\begin{eqnarray}
\hat{\Delta}^{RD} &=& \frac{\sum_{k=1}^K (Y_k-n_k \pi_{k,0})}{\sum_{k=1}^K w_k n_k } = \frac{\sum_{k=1}^K R_k^{RD}}{\sum_{k=1}^K S_k^{RD} }, 
\label{mh_rd}
\end{eqnarray}
where $R_k^{RD}=Y_k-n_k \pi_{k,0}$ and $S_k^{RD}=n_k$. The construction of the estimator is same as the idea of the standard Mantel-Haenszle estimator for stratified contingency tables (Greenland and Robins 1985). Then, we call the estimator $\hat{\Delta}^{RD}$ the one-sample Mantel-Haenszel risk difference (MH-RD) estimator. 
Note that from the definition ($\ref{mh_rd}$), one can see that the estimator $\hat{\Delta}^{RD}$ is regarded as the solution to the estimating equation $U^{RD}(\Delta^{RD})=0$, where
\begin{eqnarray*}
U^{RD}(\Delta^{RD}) &=& 
\sum_{k=1}^K R_k^{RD}-\Delta^{DR}  \sum_{k=1}^K S_k^{RD}.
\label{eqrd1}
\end{eqnarray*}
Similarly to the standard Mantel-Haenszel estimators for stratified contingency tables, we consider asymptotic justification under the two limiting models (Breslow 1981); \\
Asymptotic 1 (the large strata limiting model): $K$ is fixed and $n_k \to \infty$ and $n_k/n \to q_k \in (0,1)$, \\
and \\
Asymptotic 2 (the sparse strata limiting model): $K \to \infty$ and $n_k$ is fixed and bound over $k=1,2,...,K$. \\ \\
It is shown that MH-RD estimator is dually consistent if the common RD assumption ($\ref{rd1}$) holds. That is, MH-RD converges in probability to $\Delta^{RD}$ if either of Asymptotic 1 or Asymptotic 2 holds (see Appendix A-1 and A-2). 
From the relationship $\sqrt{K}(\hat{\Delta}^{RD}-\Delta) \simeq \{K^{-1}dU^{RD}(\Delta^{RD})/d\Delta^{RD}\}^{-1} K^{-\frac{1}{2}}U^{RD}(\Delta^{RD})$, the following variance estimator
\begin{eqnarray}
V\hat{a}r(\hat{\Delta}_k^{RD} ) &=& 
\frac{\sum_{k=1}^K \frac{n_k^2}{n_k-1} \hat{\pi}_k(1-\hat{\pi}_k)}
{(\sum_{k=1}^K S_k^{RD})^2}.
\label{v2_rd}
\end{eqnarray}
is motivated. 
The asymptotic normality also holds in the dual sense and its asymptotic variance can be dually consistently estimated by $V\hat{a}r(\hat{\Delta}^{RD} )$ in ($\ref{v2_rd}$). A proof is given in Appendix A-3. 
Then, a two-tailed $100(1-\alpha)$ percent confidence interval of $\Delta^{RD}$ is given by $\hat{\Delta}^{RD} \pm z_{1-\alpha/2} \sqrt{V\hat{a}r(\hat{\Delta}^{RD})}$.

\subsection{Mantel-Haenszel risk ratio}
Denote the RR of the $k$th strata by $\Delta_k^{RR}=\pi_k/\pi_{k,0}$. 
In this subsection, we introduce inference under the common RR assumption; for $k=1,2,...,K$, 
\begin{eqnarray}
\Delta^{RR}=\Delta_k^{RR}=\frac{\pi_k}{\pi_{k,0}}.
\label{rr1}
\end{eqnarray}
Replacing $\pi_k$ in ($\ref{rr1}$) with $\hat{\pi}_k$, $\Delta^{RR} n_k \pi_k \simeq Y_k$ and then
\begin{eqnarray}
\Delta^{RR} \sum_{k=1}^K w_k n_k \pi_k \simeq \sum_{k=1}^n w_k Y_k
\label{rr1b}
\end{eqnarray} 
are suggested, where $w_k$ is a known weight, which is positive and uniformly bounded over $k=1,2,...,K$. For the RR, we introduce the weight $w_k$ to make the estimator interpretable under misspecification. See Section 4.3. The relationship ($\ref{rr1b}$) motivates the one-sample Mantel-Haenszel risk ratio (MH-RR) estimator;
\begin{eqnarray*}
\hat{\Delta}^{RR} &=& \frac{\sum_{k=1}^K w_k Y_k}{\sum_{k=1}^K w_k n_k \pi_{k,0}} =  \frac{\sum_{k=1}^K R_k^{RR}}{\sum_{k=1}^K S_k^{RR}} 
\label{mh_rr}
\end{eqnarray*}
where $R_k^{RR}=w_k n_k $ and $S_k^{RR}=w_k n_k \pi_k$. 

Similarly to the case of the RD, the MH-RR estimator $\hat{\Delta}^{RR}$ is regarded as the solution to the estimating equation $U^{RR}(\Delta^{RR})=0$, where
\begin{eqnarray*}
U^{RR}(\Delta^{RR}) &=& 
\sum_{k=1}^K R_k^{RR}-\Delta^{RR} \sum_{k=1}^K S_k^{RR}.
\label{eqrr1}
\end{eqnarray*}
From this estimating equation representation, one has a variance estimator
\begin{eqnarray}
V\hat{a}r(\hat{\Delta}_k^{RR} ) &=& 
\frac{\sum_{k=1}^K \frac{n_k^2}{n_k-1} w_k^2 \hat{\pi}_k(1-\hat{\pi}_k)}
{(\sum_{k=1}^K S_k^{RR})^2}.
\label{v2_rr}
\end{eqnarray}
Similarly to the case of the RD, as presented in Appendix B, consistency and  asymptotic normality of $\hat{\Delta}^{RR}$, and consistency of $V\hat{a}r(\hat{\Delta}^{RR})$ can be shown under Asymptotic A. Under Asymptotic 2, coupled with some additional regularity conditions, they are also shown in Appendix B.

\subsection{Interpretation when the common parameter assumption is violated}
Both the MH-RD and MH-RR estimators assumed the common RD ($\ref{rd1}$) and RR ($\ref{rr1}$) assumptions, respectively. In this subsection, asymptotic properties of the MH-RD and MH-RR estimators when these common parameter assumptions are violated. This problem was considered by Noma and Nagashima (2016) for the standard Mantel-Haenszel etimators for stratified $2 \times 2$ tables. We begin with the MH-RD estimator. As presented in Appendix A-1, under Asymptotic 1, it holds
\begin{eqnarray*}
\hat{\Delta}^{RD} \overset{p}{\longrightarrow} \frac{\sum_{k=1}^K q_k \Delta_k^{RD}}{\sum_{k=1}^K q_k},
\end{eqnarray*}
where $\overset{p}{\longrightarrow}$ shows convergence in probability. Under Asymptotic 2, it is shown in Appendix A-1 that if the limit of $K^{-1} \sum_{k=1}^K n_k \Delta_k^{RD}$ exists as $K \to \infty$, it holds that
\begin{eqnarray*}
\hat{\Delta}^{RD} \overset{p}{\longrightarrow} \frac{\lim_{K \to \infty} K^{-1} \sum_{k=1}^K n_k \Delta_k^{RD}}
{\lim_{K \to \infty}  K^{-1} \sum_{k=1}^K n_k}.
\end{eqnarray*}
In both cases, the limit is regarded as the average of the basket-specific RD weighted by sample size .
That is, the MH-RD estimator is regarded as the common RD if the common RD assumption  ($\ref{rd1}$) holds and even if violated, is still able to be interpreted as the average of the basket-specific RDs weighted by sample size. This always-interpretable feature is attractive since it is difficult to verify the assumptions statistical methods rely on, in particular at the design stage.  

Next, we examine the MH-RR estimator. Under Asymptotic 1, it holds that
\begin{eqnarray*}
\hat{\Delta}^{RR} \overset{p}{\longrightarrow} \frac{\sum_{k=1}^K w_k q_k \pi_{k,0} \Delta_k^{RR}}{\sum_{k=1}^K w_k q_k \pi_{k,0}}.
\end{eqnarray*}
Under Asymptotic 2, coupled with some regurality conditions, it is shown in Appendix B-1 that if the limit of $K^{-1} \sum_{k=1}^K w_k n_k \Delta_k^{RD}$ exists as $K \to \infty$, it holds that
\begin{eqnarray*}
\hat{\Delta}^{RR} \overset{p}{\longrightarrow} \frac{\lim_{K \to \infty} K^{-1} \sum_{k=1}^K w_k n_k \pi_{k,0} \Delta_k^{RR}}
{\lim_{K \to \infty} K^{-1} \sum_{k=1}^K w_k n_k \pi_{k,0} }.
\end{eqnarray*}
In both cases, the limit can be regarded as the weighted average of the basket-specific RR, but not by sample size. We set $w_k=\pi_{k,0}^{-1}$. Then, the resulting MH-RR estimator is inversely weighted by the clinically relevant minimum response rate. We call it the inverse probability weighted MH-RR (MH-iwRR) estimator and denoted by $\hat{\Delta}^{iwRR}$. The MH-iwRR estimator is interpreted as the common RR if the common RR assumption ($\ref{rr1}$) holds and as the average of the basket-specific RR weighted by sample size. 

In the standard Mantel-Haenszel procedure for the stratified $2\times 2$ tables, the common parameter assumption is often addressed with the Brewlow-Day test (Breslow and Day 1980). For the one-sample case, we can apply the standard chi-squared goodness-of-fit test (see Appendix C). It would be helpful in discussing heterogeneity among baskets.

\section{Classification of baskets into homogeneous subclasses via the generalized information criterion}
The Akaike information criterion (AIC) is a useful model-selection criterion, which is widely used (Akaike 1974). In its construction, discrepancy between the fitted model and the true distribution is measured by the Kullback-Leibler divergence. The AIC is regarded as the bias-corrected estimate of the Kullback-Leibler divergence. The AIC was designed for the maximum likelihood estimators. The Generalized information criterion (GIC) is an extension of the AIC, which allows estimators other than the maximum likelihood estimators (Konishi and Kitagawa 1996). The GIC covers estimators represented by a statistical functional and then allows to evaluate statistical models estimated with estimating equations (M-estimators).

Here, we treat the MH-RD and the MH-RR in a unified way. Let $\Delta$ is $\Delta^{RD}$ and $\Delta^{RR}$ when the MH-RD and the MH-RR are considered, respectively, and $\hat{\Delta}$ is defined in a similar way.  Similarly, $R_k$, $S_k$ and $U(\Delta)$ are defined. Define $h_k(x)=\pi_{k,0}+x$ when the MH-RD is considered and $h_k(x)=\pi_{k,0}x$ when the MH-RR is considered. Then, $\pi_k=h_k(\Delta)$ holds both for the MH-RD and the MH-RR. Recall that $\hat{\Delta}$ is regarded as the solution to 
\begin{eqnarray}
U(\Delta)=\sum_{k=1}^K (R_k-\Delta S_k)=0.
\label{eq1}
\end{eqnarray}
Applying the general theory for M-estimators by Konishi and Kitagawa (1996), the GIC was given by
\begin{eqnarray}
GIC=-\sum_{k=1}^K \Big[ Y_k log h_k(\hat{\Delta})+(n_k-Y_k)log(1-h_k(\hat{\Delta}))\Big]+bias(\hat{\Delta}),
\label{gic}
\end{eqnarray}
where 
\begin{eqnarray}
bias(\Delta) = (\sum_{k=1}^K S_k)^{-1} \sum_{k=1}^K (R_k-\Delta S_k)
\Big\{
Y_k \frac{\dot{h}_k(\Delta)}{h_k(\Delta)}-(n_k-Y_k)\frac{\dot{h}_k(\Delta)}{1-h_k(\Delta)}
\Big\}
\label{bias}
\end{eqnarray}
and $\dot{h}_k(x)=dh_k(x)/dx$. It is regarded that the smaller the $GIC$ is, the better the model fits. The first term of the $GIC$ comes from minus log-likelihood. The better the model fits, the smaller the first term of the $GIC$ is. That is, if we separate the baskets more, the first term is decreased. In particulate, if we separate all the baskets, the first term is minimized. The bias term acts as the penalty to increase model complexity. Thus, the GIC is expected to select the classification of baskets having a good balance between simplicity and fit of the model. 
In Appendix D-1, we show that the GIC ($\ref{gic}$) is constructed with dually consistent bias-correction. 

We consider evaluating whether baskets should be separated into several subclasses of homogeneous efficacy or not. Suppose there are $L$ subclasses of baskets, within each of which the RD (or RR) is homogeneous. For simplicity, we explain with the case of $L=2$. Extension to the general cases of $L \ge 2$ is straightforward. $K$ baskets are divided into two subclasses, denoted by $B_1$ and $B_2$. Suppose $K_1$ and $K_2$ baskets belong to $B_1$ and $B_2$, respectively, and without loss of generality, the first $K_1$ baskets are in $B_1$ and the other $K_2$ baskets are in $B_2$. Assuming the common RD (or RR) assumption within each subclass, the $GIC$ is calculated. The $GIC$s for $B_1$ and $B_2$ are denoted by $GIC_1$ and $GIC_2$. Then, the $GIC$ for the entire baskets are calculated by $GIC=GIC_1+GIC_2$. By comparing the $GIC$s among candidate classifications, one can have some insights about which baskets the treatment is effective for. In Appendix D-2, we give an intuitive explanation on effectiveness of the GIC in a simple and ideal situation.

\section{Examples}
\subsection{Vemurafenib data for BRAF-positive non-melanoma cancer}
Hyman {\it et al.} (2015) reported results of a basket trial for vemurafenib in BRAF V600 mutation-posivite non-meranoma cancers. Six cancer types were predefined as baskets; anaplastic thyroid cancer (ATC), Erdheim- Chester disease of Langerhans'-cell histocytosis (ECD/LCH)), cholangiocarcinoma (CCA), colorectal cancer treated by vemurafenib (CRC-V), colorectal cancer treated by vemurafenib and cetuximab (CRC-VC), and non-small cell lung cancer (NSCLC). The Simon's two-stage method was used separately for each basket with the tumor response at week 8 as the primary endpoint. The clinically meaningful minimum response rate was defines as $\pi_{k,0}=0.15$ for all the baskets. The number of patients $n_{k}$ and that of responders $Y_k$ in each basket are listed, as well as the response rate with Pearson-Cropper two-tailed 95 percent confidence intervals, in Table 1. The data was re-analyzed by Hobbs and Landin (2018) and Zhou and Ji (2020) in their Bayesian ways, respectively. A summary of the results with these methods is given in Table 4 of Zhou and Ji (2020). See Hobbs and Landin (2018) and Zhou and Ji (2020)  for details of specifications. Hobbs and Landin (2018) showed that the posterior probability $P(\pi_k \ge \pi_{k,0}|data)$ was 0.97, 1, and 1 for the baskets 1 (ATC), 2 (ECD/LCH) and 6 (NSCLC), respectively, and those were very low for other baskets. As presented in Zhou and Ji (2020), the baskets 2 and 6 had very evident Bayes factors 124 and 332, respectively, whereas the Bayes factor of the basket 1 was 3.2, which was not substantially evident against $\pi_{k,0}=0.15$. Zhou and Ji (2020) also reported that the posterior mode of the number of the latent subgroups was 3 with the posterior probability 0.34. The posterior probability for the 2 latent subgroups were 0.32, close to the posterior mode.    

We illustrate our proposed method with this data contrasting to these Bayesian results. The MH-RD and MH-iwRR estimators, as well as basket-specific ones, are presented in Table 1. For the RD, we used the test statistic $T=\sum_{k=1}^K Y_k$. The one-tailed P-value of the exact test was 0.0710, which was calculated with 10,000 simulated samples from the null distribution. The MH-RD estimate was 0.064 (95 percent CI: -0.017, 0.146). The confidence interval contained the null value of 0. For the RR, we used the weight $w_k=\pi_{k,0}^{-1}$. The resulting test statistic is denoted by $T_{iw}=\sum_{k=1}^K \pi_{k,0}^{-1} Y_k$, which gave the P-value 0.0707.  The MH-RR estimate was 1.43 (0.884, 1.972). These results also failed to establish the effectiveness of the treatment at the two-tailed 5 percent significance level. However, these small P-values might suggest effectiveness for some baskets. 

From the basket-specific RD estimates, heterogeneity of the RD  (or the RR) was suggested. Indeed, the chi-squared goodness-fit-test gave a P-value of 0.022. We calculated the GICs of the all the possible two subclassifications, as well as the GIC under the common parameter assumption. For this example, we focus on the RD. There were 31 models with two classes of baskets.  Then, with the model under common RD assumption,  we compared totally 32 models with the GIC. In Table 2, the 5 top and the last 5 models with respect to GICs are listed. The top model with the minimum GIC consisted of a subclass of the baskets 1, 2 and 6 and that of the baskets 3, 4, 5. Here, we denote the subclass consisting of the basket 1,2 and 6 by $B_{1,2, 6}$. Similar notations are used for other cases. In Table 2, the MH-RD estimate of each subclass are presented; we placed the class of smaller estimate (less effective) on the first row in each model. The MH-RD for $B_{1, 2, 6}$ was 2.50 (0.93, 0.407). The 95 percent confidence interval excluded the null value and then effectiveness of the treatment for these 3 baskets was suggested. 
The model of the second smallest GIC had the separation of $B_{1, 2, 3, 6}$ and $B_{4, 5}$. The subclass $B_{4, 5}$ had a negative RD. The subclass of the third smallest GIC had the separation of $B_{2, 6}$ and $B_{1, 3, 4, 5}$. The basket-specific RD estimates and the Bayesian consideration summarized above consistently suggested substantial effectiveness of the baskets 2 and 6. Table 2 also indicates that the models with the 5 largest GICs (the last 5 models) failed to construct two subclasses of good separation of the RD. It suggests that the proposed GIC could work to identify subclasses of good effectiveness. 

We also evaluate the model of three subclassification of $B_{1, 3}$, $B_{2, 6}$ and $B_{4, 5}$. Its GIC was 35.029, which was slightly smaller than but almost same as the top model of $B_{1, 2, 6}$ and $B_{3, 4, 5}$. This is consistent with the observations by Zhou and Ji (2020) on the posterior mode of the number of latent subgroups (see the end of the first paragraph of Section 6). 
The GIC solely can not conclude whether the basket 1 should be regarded as similarly effective to the baskets 2 and 6.

\subsection{Imatinib data for advanced sarcoma}
Chugh et al. (2009) reported results of a basket study for imatinib in patients with advanced sarcoma. It allowed to enroll 10 subtypes of sarcoma. The primary endpoint was the clinical benefit rate (CBR), which was defined as Complete Response, Partial Response within 16 week or lasting Stable Disease at least 16 weeks. A Bayesian hierarchical model (BHM; Thall et al. 2003) was predefined for the primary analysis. The clinically meaningful minimum response rate was set as 0.3 throughout the baskets and basket-specific efficacy was adaptively assessed by monitoring $P(\pi_{k}>0.3|data)$ for $k=1,2,...,10$ borrowing information across baskets. For all the baskets, clinically meaningful efficacy was not established. We use this study for illustration modifiying the setting. We set the clinically meaningful minimum response rate was set as $\pi_{k,0}=0.1$ for all the baskets. In Table 3, the number of patients $n_{k}$, that of responders $Y_k$ and the response rate with Pearson-Cropper two-tailed 95 percent confidence intervals are summarized. The one-sided P-value of the exact test was 0.0120 both with $T$ and $T_{iw}$. The MH-RD and MH-iwRR estimates was presented as 0.056 (0.003, 0.110) and 1.564 (1.029, 2.100), respectively. Both confidence intervals indicated statistical significance at the two-tailed 5 percent significance level, which were coherent with the exact text. The exact test and the Mantel-Haenszel estimator suggested that Imatinib is effective at least for some subtypes of sarcoma. The chi-squared goodness-of-fit test gave p=0.784, which did not suggest substantial violation of the common RR assumption.   

An interesting feature of this data was there were two baskets with very small number of patients; the baskets 8 and 9 had only 5 and 2 patients, respectively. As summarized in Table 3 of Zhou and Ji (2020), for the basket 8 (MPNST), the BHM by Thall et al. (2003), EXNEX by Neuenschwander et al. (2016) and RoBoT by Zhou and Ji (2020) provided very similar posterior mean for the response rates around 0.15 with 95 percent credible intervals around (0.01, 0.50). For the basket 9 (Rhabdomyosarcoma), BHM and EXNEX gave the posterior mean as 0.101 (95 $\%$ credible interval: 0.1, 0.509) and 0.131 (0.001, 0.429), respectively, whereas RoBoT provided a much smaller posterior mean 0.003 (0.000, 0.022). 
Hereafter, we focus on RR. In Table 4, we listed the top 10 models with respect to the GIC. As given in Table 4, the best model had the minimum GIC of 79.663 and gave the separation $B_{1,2,3,6,9,10}$ and $B_{4, 5, 7, 8}$. The MH-iwRR estimate for $B_{4,5,7,8}$ was 2.159 (1.280, 3.038), whereas it was 0.989 (0.367, 1.611) for $B_{1,2,3,6,9,10}$. The GIC seems to separate effective and ineffective baskets reasonably. The basekts 7 and 8 had a high basket-specific RR with large sample sizes. They were classified into the subclass of effective baskets in all the top 10 models. The baskets 2, 3 and 6 had a low basket-specific RR with more than 10 subjects. These baskets were classified into the subclass of ineffective baskets in almost all the top 10 models. These observation suggested that the GIC would be a useful guide to evaluate effectiveness of baskets with certainly large number of patients. According to the best model, the basket 8 was regarded as a basket of effectiveness to the treatment and the basket 9 was as a basket of ineffectiveness to the treatment. On the other hand, since the number of patients in these baskets was very small. Then, instability of the result is concerned. Among 9 of them had GICs within 1 from the minimum GIC of 79.663, which was corresponding to statistically significant difference from the top model. Note that the upper column in each model shows the subclass of less effectiveness. We highlighted the baskets 8 and 9 with bold. Among the top ten models, the basket 8 was classified as the effective subclass in 6 models. The basket 9 was in 5 models. It indicated that the classification of these two baskets only with small sample size were very instable and then it should be noted in making any decisions on such baskets. Thus, we recommend not to stick to a single best model, but to check sub-optimal but good models.

\section{Simulation study}
\subsection{Accuracy of the common parameter estimation}  
To examine whether the proposed methods worked in practical situations, some simulation studies were carried out. We considered the cases of $K=6$ and $K=10$. For the case of $K=6$, we set $(n_1, n_2, n_3, n_4, n_5, n_6)=(7, 14, 8, 26, 19, 19)$ as the vemurafenib data listed in Table 1 and for the case of $K=10$, $(n_1, n_2, n_3, n_4, n_5, n_6, n_7, n_8, n_9, n_{10})=(15, 13, 12, 28, 29, 29, 26, 5, 2, 20)$ as the imatinib data listed in Table 3. We evaluated the MH-RD, MH-RR with $w_k=1$ and MH-iwRR estimators. We generated three kinds of datasets; one was under the null hypothesis of no treatment effects over all the baksets, which is referred as $Null$, and the other two were under the alternative hypothesis, in which the treatment was effective for all or some baskets. The first sets of datasets under the alternative hypothesis are referred as $aRD$, in which the RD were homogeneous over all the baskets or over part of baskets. The second set of datasets under the alternative hypothesis are referred as $aRR$, in which the RR were homogeneous over all the baskets or over part of baskets. 
 For each combination of $K=6$ or $K=10$ and RD or RR, we generated four kinds of datasets with combinations of low or moderate $\pi_{k,0}$ and homogeneous or heterogeneous parameters. Details of the settings are presented in Table 5. For example, the dataset RD-A-1-1 had $K=6$ baskets of low $\pi_{k,0}$ and homogeneous RDs and the dataset RD-B-2-2 had $K=10$ baskets of moderate $\pi_{k,0}$ and heterogeneous RDs. For each scenario, 10,000 datasets were generated. 

In Table 6, we summarized the results, in which biases and coverage probabilities of the MH estimators were evaluated. For the scenarios under the null hypothesis, we evaluated empirical type 1 error probabilities for the one-tailed 2.5 percent nominal level of the asymptotic Wald-type tests and the exact test. The MH estimators successfully estimated the null values. In some scenarios, the type 1 error probabilities could be inflated with the asymptotic MH methods. On the other hand, the exact test fairly controlled the type 1 error probabilities. 
With scenarios referred as $aRD$, performance of the MH-RD was examined, in which the column "True" shows the average of the basket-specific RDs weighted by the proportion of the sample size. Both when the common RD assumption held (RD-A-1-1, RD-A-1-2, RD-B-1-1, RD-B-1-2) and when not, the MH-RD estimators had only negligible biases and the dually consistent variance estimator had empirical coverage probabilities close to the nominal level of 95 percent. 
With the scenarios referred as $aRR$, we evaluated performance of the MH-RR and MH-iwRR estimators. The MH-RR estimator had only negligible biases when the homogeneous RR assumption held (RR-A-1-1, RR-A-1-2, RR-B-1-1, RR-B-1-2). When the assumption is violated, it had biases (RR-A-2-1, RR-A-2-2, RR-B-2-1, RR-B-2-2). On the other hand, the MH-iwRR estimator had only negligible biases even in such cases. However its empirical coverage probabilities might be a bit away from the nominal level of 95 percent in particular when the number of baskets was small  (K=6) (RR-A-1-1, RR-A-2-1). This tendency was weakened with K=10. (RR-B-1-1, RR-B-2-1). The inverse probability weighting could be responsible for this observation; in these cases, MH-iwRR estimator was inversely weighted by the low clinically meaningful minimum response rates. Indeed, the tendency of the poor coverage probabilities was weakened with larger $\pi_{k,0}$. (RR-A-1-2, RR-A-2-2, RR-B-1-2, RR-B-2-2).

\subsection{Performance of GIC-based basket identification}
We conducted simulation studies to evaluate usefulness of the GIC for identifying effective baskets. We generated simulation datasets following the setting by Zhou and Ji (2020), in which several Bayesian methods were compared. We supposed $K=4$ baskets of ($n_1$, $n_2$, $n_3$, $n_4$)=(20, 20, 10, 10) and $\pi_{k,0}=0.1$ for $k=1, 2, 3, 4$. Seven scenarios for ($\pi_1$, $\pi_2$, $\pi_3$, $\pi_4$) were consided, which were listed in Table 1 of Zhou and Ji (2020). For $K=4$, we had 7 models with two subclasses of homogeneous RD within each subclass. With a model under the common RD assumption, we applied totally 8 models. Here, we selected the model of the minimum GIC among 8 models. We refer this method as {\it two subclass} method.    

The RD of each basket was calculated with the subclassification according to the selected model. We judged effectiveness of each basket with the statistical significance of the subclass specific MH-RD estimate. For example, suppose the model of the minimum GIC suggested the subclassification of $B_{1, 2}$ and $B_{3, 4}$, which were the subclass of baskets 1 and 2 and those of baskets 3 and 4, respectively. The baskets 1 and 2 were judged effective if the lower bound of two-tailed 95 percent confidence interval for the MH-RD with the baskets 1 and 2 was greater than 0. For the baskets 3 and 4, a similar way was applied with the subclass specific MH-RD estimate for the baskets 3 and 4. For each scenario, 10,000 datasets were generated. 

Results of the simulation study is summarized in Table 7, in which the biases, the mean-squared errors and the proportions to be rejected ($\%$Reject) are reported.  In the scenarios $1GN$ and $2GA$, the common RD assumption held. In both scenarios, the basket-specific estimates for the response rate had only negligible biases throughout the baskets. 
In $1GN$, the the proportion to be rejected in each basket was very small; the proposed method successfully protected the false-positive.  In $2GA$, $\%$Reject was around 70 percent for baskets with $n_k=20$ (baskets 1 and 2) and that was about 60 percent for the baskets with $n_k=10$ (baskets 3 and 4). 

In the scenarios 3, 4 and 5, some baskets were not effective. The proposed method successfully reduced the proportion to select ineffective baskets and had negligible biases. The scenarios 6 and 7 handled the cases, where two sub-group classification was misspecified; models classfying into three subclasses were out of consideration in model selection. Although we observed inflated $\%Reject$, the proposed method was successful to identify effective baskets.  

Recall that the simulation datasets were generated from the same setting as Zhou and Ji (2020), and then the performance of the proposed method can be compared with the five methods investigated in Zhou and Ji (2020); the scenario numbers in Table 7 are corresponding to those in Table 2 of Zhou and Ji (2020). The paper by Zhou and Ji (2020) examined five Bayesian methods; $Stratified$, in which conjugate beta-prior distributions were updated separately by baskets, Bayesian Hierarchical Model ($BHM$) by Berry {\it et al.} (2013), $EXNEX$ by Neuenschwander {\it et al.} (2016), $BLAST$ by Chu and Yuan (2018) and $RoBot$ by Zhou and Ji (2020). See Section 3.1 of  Zhou and Ji (2020) for the settings of parameters in each model. No single method dominated the others uniformly over the scenarios. It was also true for our proposed method. The proposed method dominated $Stratified$ in all the scenarios. Overall, the performance of our proposed method were very comparable to the five Bayesian methods although each method had its pros and cons depending on the scenarios. The $EXNEX$, $RoBOT$ and our method successfully protect false positive, whereas $BHM$ and $BLAST$ were likely to fail.  Recall that all the calculations required in our method can be made with a simple formula of closed-form expression. Thus, comparable performance of our proposed method to sophisticated Bayesian methods is surprising.

Recall that we selected the best model among candidate models of at most two subclasses of baskets. To see impacts of the choice of the candidates models among which the best model was selected, we also evaluated two more methods. We considered to select the best one among all the possible subclassifications. That is, in addition to 1 model under the common RD assumption and 7 models with two subclasses, 6 models of three subclasses and 1 model of 4 completely separate baskets were also considered as candidate models. We refer this method as {\it all subclasses} method. To avoid unstable estimates, we also considered to select a model avoiding models with baskets of small number of patients. We call this method as {\it non-sparse} method. Here, we avoid baskets with only 10 patients. In other word, the classification like baskets 1,2 3 and basket 4 is excluded from consideration since the MH-RD estimator for the subclass of the basket 4 relied only on 10 patients. The results with these two methods are also summarized in Table 5. Since more precise subclassification was under consideration in {\it all subclasses} method, we could avoid substantial infration of $\%$REJECT in the scenarios 6 and 7. On the other hand, it was likely to give very conservative results. Use of unstable estimates in candiates models might lead to poor performances. To address this concern, we evaluated the performance of the {\it non-sparse} method. It had comparable performance with {\it two subclasses} method overall and outperformed in some scenarios.  For example, it successfully avoided infration of $\%$REJECT in the scenrios 6 and 7 since the corresponding division of basktes were under consideration with candidate models. On the other hand, it was poorer in some scenarios, in particular the scenrio 2. In general, the {\it two subclasses} method is recommended. If subclassfication into three or more subclasses are suggested by basket-specific estimates of the response rates, additional evaluation of the GICs for models with more than 2 subclasses would be recommended. Avoiding subclasses of too small number of patients would be a good guide to consider candidate models.

\section{Discussion}
Based on the development in this paper, we propose the following strategy for the basket trial;\\
(i) The primary analysis is conducted by the exact test with $T_w=\sum_{k=1}^K w_k Y_k$ for the null hypothesis ($\ref{null}$). \\
(ii) The treatment effect is summarized with the MH-RD or the MH-iwRR. \\
(iii) The subclass of effective baskets are identified in an exploratory matter using the GIC.  \\ 
From the argument in Section 4.3, in the step (i), we recommend the constant weight $w_k=1$ for the RD and the weight $w_k=\pi_{k,0}^{-1}$ for the RR accounting for the correspondence to the MH estimators in the step (ii). 

The proposed strategy consists of two parts; one is (i) and (ii), which is the confirmatory part, and the other is (iii), an exploratory part. Although we recommend to use the GIC, we really appreciate usefulness of Bayesian methods in basket trials, in particular in the step (iii). We  recommend to conduct various analyses for (iii) from different viewpoints and comprehensive consideration is made. We believe that our confirmatory analysis based on prespecified statical analysis would be a sound foundation in interpreting results of basket trials even if they have strong exploratory aspects in addressing which baskets the treatment would be effective for.  

Most existing methods for baskets trials include steps to drop ineffective baskets at interim analysis (Li {\it et al.} 2019; Cunanan {\it et al.} 2017; Simon {\it et al.} 2016 among others). Such adaptation would be very important for basket trials since the treatment may not be effective for some baskets and stopping recruitment of patients to such baskets would be ethically sound. In this paper, we did not consider such adaptive selection. We proposed to use the exact test by London and Chang (2005) as the primary analysis. 
London and Chang (2005) proposed two-stage procedure. We can use their method. Furthermore, we may introduce adaptive procedure to drop ineffective baskets using the GIC, which would be a valuableresearch topic.  

In this paper, we focus on the RD and the RR since they are easier to interpret than the odds ratio. As summarized in Appendix E, in a similar argument, one can consider the one-sample Mantel-Haenszel estimator for the common odds ratio.

\section*{Acknowlegements}
\label{s:acknow}
The first author's research was partly supported by Grant-in-Aid for Scientific Research(18H03208) from the Ministry of Education, Science, Sports and Technology of Japan. \vspace*{-8pt}

\section*{Data availability statement}
We used a dataset available to the public.

\newpage
\clearpage

\begin{table}[]
\caption[]
	{\textit{Vemurafenib data for BRAF-positive non-melanoma cancer; in the blaket for the response rate, two-tailed 95 percent Cropper-Peason confidence interval is show. 
	}}
\begin{center}
\begin{tabular}{lcccccc}
\hline 
basket & $Y_k$ & $n_k$  & $\hat{\pi}_{k} (95\%CI)$   & $\pi_{k,0}$    & $RD$     & $RR$    \\
\hline
1: ATC        & 2 & 7  & 0.286 (0.037, 0.710)         & 0.150 & 0.136           & 1.905          \\
2: ECD/LCH       & 6 & 14 & 0.429 (0.177, 0.711)         & 0.150 & 0.279           & 2.857          \\
3: CCA     & 1 & 8  & 0.125 (0.003, 0.527)         & 0.150 & -0.025          & 0.833          \\
4: CRC-V       & 1 & 26 & 0.038 (0.001, 0.196)         & 0.150 & -0.112          & 0.256          \\
5: CRC-VC       & 0 & 10 & 0.000 (0.000, 0.309)        & 0.150 & -0.150          & 0.000          \\
6: NSCLC        & 8 & 19 & 0.421 (0.209, 0.665)         & 0.150 & 0.271           & 2.807          \\
\hline
Mantel-Haenszel      &   &    &               &       & 0.064           & 1.429          \\
(95\%CI) &   &    &               &       & (-0.017, 0.146) & (0.884, 1.973) \\
Exact test &   &    &               &       & P=0.0710 & P=0.0707 \\
\hline
\end{tabular}
\end{center}
\end{table}

\begin{table}[]
\caption[]
	{\textit{Sub-class specific RD with top or last 5 models with the GIC, as well as estimates with the common RD assumption for Vemurafenib data.  
	}}
\begin{center}
\begin{tabular}{cccclc}
\hline
       & Rank & GIC    & Model     & Sub-class    & MH-RD (95$\%$CI)             \\
\hline
Top5   & 1    & 35.494 & 1 2 6/ 3 4 5 & 3 4 5       & -0.105 ( -0.168, -0.042) \\
          &      &        &              & 1 2 6       & \ 0.250 ( \ 0.093,  \ 0.407)  \\
          & 2    & 36.501 & 1 2 3 6/ 4 5 & 4 5         & -0.122 ( -0.177, -0.068) \\
          &      &        &              & 1 2 3 6     & \ 0.204 ( \ 0.067, \ 0.341)  \\
          & 3    & 37.584 & 1 3 4 5/ 2 6 & 1 3 4 5     & -0.072 ( -0.145, \ 0.002) \\
          &      &        &              & 2 6         & \ 0.274 ( \ 0.100, \ 0.448)  \\
          & 4    & 39.623 & 1 4 5/ 2 3 6 & 1 4 5       & -0.080 ( -0.155, -0.006) \\
          &      &        &              & 2 3 6       & \ 0.216 ( \ 0.068, \  0.364)  \\
          & 5    & 40.745 & 1 2 5 6/ 3 4 & 3 4         & -0.091 ( -0.173, -0.010) \\
          &      &        &              & 1 2 5 6     & \ 0.170 ( \ 0.045, \ 0.295)  \\
          &      &        &              &             &                         \\
homogeneous      & 17   & 47.228 & 1 2 3 4 5 6  & 1 2 3 4 5 6 & 0.064 ( -0.017, \ 0.146)  \\
          &      &        &              &             &                         \\
Last 5 & 28   & 49.823 & 1 3 5 6/ 2 4 & 2 4         & 0.025 ( -0.081, \ 0.131)  \\
          &      &        &              & 1 3 5 6     & 0.100 ( -0.022, \ 0.222)  \\
          & 29   & 49.906 & 1 2 4/ 3 5 6 & 1 2 4       & 0.041 ( -0.064, \ 0.147)  \\
          &      &        &              & 3 5 6       & 0.093 ( -0.035, \ 0.222)  \\
          & 30   & 50.048 & 1 3 4 6/ 2 5 & 1 3 4 6     & 0.050 ( -0.046, \ 0.146)  \\
          &      &        &              & 2 5         & 0.100 ( -0.057, \ 0.257)  \\
          & 31   & 50.390 & 1 4 6/ 2 3 5 & 1 4 6       & 0.062 ( -0.042, \ 0.165)  \\
          &      &        &              & 2 3 5       & 0.069 ( -0.064, \ 0.201)  \\
          & 32   & 50.488 & 1 2 3 5/ 4 6 & 4 6         & 0.050 ( -0.056, \ 0.156)  \\
          &      &        &              & 1 2 3 5     & 0.081 ( -0.046, \ 0.208)  \\
\hline
\end{tabular}
\end{center}
\end{table}

\begin{table}[]
\caption[]
	{\textit{Imatinib data for advanced sarcoma
	}}
\begin{center}
\begin{tabular}{lccccccc}
\hline 
basket & $Y_k$ & $n_k$  & $\hat{\pi}_{k} (95\%CI)$   & $\pi_{k,0}$    & $RD$     & $RR$    \\
\hline
1:Angiosarcoma      & 2 & 15 & 0.133 (0.017, 0.405) & 0.100 & 0.033  & 1.333 \\
2:Ewing      & 0 & 13 & 0.000 (0.000, 0.258) & 0.100 & -0.100 & 0.000 \\
3:Fibrosarcoma      & 1 & 12 & 0.083 (0.002, 0.385) & 0.100 & -0.017 & 0.833 \\
4:Leiomyosarcoma      & 6 & 28 & 0.214 (0.083, 0.410) & 0.100 & 0.114  & 2.143 \\
5:Liposarcoma      & 7 & 29 & 0.241 (0.103, 0.435) & 0.100 & 0.141  & 2.414 \\
6:MFH      & 3 & 29 & 0.103 (0.022, 0.274) & 0.100 & 0.003  & 1.034 \\
7:Osteosarcoma      & 5 & 26 & 0.192 (0.066, 0.394) & 0.100 & 0.092  & 1.923 \\
8:MPNST      & 1 & 5  & 0.200 (0.005, 0.716) & 0.100 & 0.100  & 2.000 \\
9:Rhabdomyosarcoma      & 0 & 2  & 0.000 (0.000, 0.842) & 0.100 & -0.100 & 0.000 \\
10:Synovial     & 3 & 20 & 0.150 (0.032, 0.379) & 0.100 & 0.050  & 1.500 \\
\hline
Mantel-Haenszel        &   &    &       &       & 0.056          & 1.564          \\
(95\%CI) &   &    &       &       & (0.003, 0.110) & (1.029, 2.100) \\
Exact test &   &    &       &       & P=0.0120 & P=0.0120 \\
\hline
\end{tabular}
\end{center}
\end{table}

\begin{table}[]
\caption[]
	{\textit{Sub-class specific iwRR with top 10 models with the GIC for Imatinib data. 
	}}
\begin{center}
\begin{tabular}{ccclc}
\hline
Rank & GIC    & Model                 & Sub-class         & MH-iwRR (95$\%$CI)                   \\
\hline
1    & 79.663 & 1 2 3 6 9 10/ 4 5 7 8 & 1 2 3 6 {\bf 9} 10       & 0.989 (  0.367,  1.611) \\
     &        &                       & 4 5 7 {\bf 8}            & 2.159 (  1.280,  3.038) \\
2    & 79.705 & 1 2 3 6 8 9 10/ 4 5 7 & 1 2 3 6  {\bf 8}  {\bf 9} 10     & 1.042 (  0.418,  1.665) \\
     &        &                       & 4 5 7              & 2.169 (  1.267,  3.071) \\
3    & 79.871 & 1 3 4 5 6 7 8 10/ 2 9 & 2 {\bf 9}                & 0.000 (  0.000,  0.000) \\
     &        &                       & 1 3 4 5 6 7  {\bf 8}  10   & 1.707 (  1.123,  2.292) \\
4    & 80.007 & 1 2 3 6 10/ 4 5 7 8 9 & 1 2 3 6 10         & 1.011 (  0.376,  1.647) \\
     &        &                       & 4 5 7  {\bf 8}  {\bf 9}          & 2.111 (  1.251,  2.971) \\
5    & 80.083 & 1 2 3 6 8 10/ 4 5 7 9 & 1 2 3 6  {\bf 8}  10       & 1.064 (  0.427,  1.701) \\
     &        &                       & 4 5 7 {\bf 9}            & 2.118 (  1.237,  2.998) \\
6    & 80.328 & 1 3 4 5 6 7 8 9 10/ 2 & 2                  & 0.000 (  0.000,  0.000) \\
     &        &                       & 1 3 4 5 6 7  {\bf 8} {\bf 9} 10 & 1.687 (  1.109,  2.264) \\
7    & 80.374 & 1 2 3 6 9/ 4 5 7 8 10 & 1 2 3 6 {\bf 9}          & 0.845 (  0.189,  1.501) \\
     &        &                       & 4 5 7  {\bf 8}  10         & 2.037 (  1.261,  2.813) \\
8    & 80.516 & 1 2 3 6 8 9/ 4 5 7 10 & 1 2 3 6  {\bf 8}  {\bf 9}        & 0.921 (  0.256,  1.586) \\
     &        &                       & 4 5 7 10           & 2.039 (  1.248,  2.830) \\
9    & 80.657 & 1 2 3 6/ 4 5 7 8 9 10 & 1 2 3 6            & 0.870 (  0.195,  1.544) \\
     &        &                       & 4 5 7  {\bf 8}  {\bf 9} 10       & 2.000 (  1.238,  2.762) \\
10   & 80.864 & 1 2 3 6 8/ 4 5 7 9 10 & 1 2 3 6  {\bf 8}           & 0.946 (  0.263,  1.629) \\
     &        &                       & 4 5 7 {\bf 9} 10         & 2.000 (  1.224,  2.776) \\
\hline
\end{tabular}
\end{center}
\end{table}

\begin{landscape}
\begin{table}[]
\caption[]
	{\textit{Summary of the simulation design; $(n_1, n_2, n_3, n_4, n_5, n_6)=(7, 14, 8, 26, 10, 19)$ for $K=6$ and
 $(n_1, n_2, n_3, n_4, n_5, n_6, n_7, n_8, n_9, n_{10})=(15, 13, 12, 28, 29, 29, 26, 5, 2, 20)$ for $K=10$
	}}
\begin{center}
\small
\begin{tabular}{lllll}
\hline
 & $K$ & dataset    & ($\pi_{1,0}, \cdots , \pi_{k,0}$)                                             & RD or RR                       \\
\hline
{\it Null}    & 6             & Null-A-1-1 & (0.15, 0.15, 0.1, 0.1, 0.05, 0.05)                      & RD=0 or RR=1 for all baskets                       \\
        &               & Null-A-1-2 & (0.35, 0.35, 0.3, 0.3, 0.2, 0.2)                        & RD=0 or RR=1 for all baskets                       \\
        & 10            & Null-B-1-1 & (0.15, 0.15, 0.1, 0.1, 0.05, 0.05)                      & RD=0 or RR=1 for all baskets                       \\
        &               & Null-B-1-2 & (0.35, 0.35, 0.3, 0.3, 0.2, 0.2)                        & RD=0 or RR=1 for all baskets                       \\
        &               &            &                                                         &                                                    \\
{\it aRD}      & 6             & RD-A-1-1   & (0.15, 0.15, 0.1, 0.1, 0.05, 0.05)                      & (0.1, 0.1, 0.1, 0.1, 0.1, 0.1)                     \\
        &               & RD-A-1-2   & (0.35, 0.35, 0.3, 0.3, 0.2, 0.2)                        & (0.1, 0.1, 0.1, 0.1, 0.1, 0.1)                     \\
        &               & RD-A-2-1   & (0.15, 0.15, 0.1, 0.1, 0.05, 0.05)                      & (0.2, 0.2, 0.2, 0, 0, 0)                           \\
        &               & RD-A-2-2   & (0.35, 0.35, 0.3, 0.3, 0.2, 0.2)                        & (0.2, 0.2, 0.2, 0, 0, 0)                           \\
        &               &            &                                                         &                                                    \\
        & 10            & RD-B-1-1   &  0.1 ($k=1,..,5$), 0.05 ($k=6,..,10$)  & RD=0.1 for all baskets \\
        &               & RD-B-1-2   & 0.3 ($k=1,..,5$), 0.25 ($k=6,..,10$) & RD=0.1 for all baskets \\
        &               & RD-B-2-1   &  0.1 ($k=1,..,5$), 0.05 ($k=6,..,10$)  & RD=0.2 ($k=1, 2, 3$), =0.1 ($k=4, 5, 6$), =0($k=7,..,10$)         \\
        &               & RD-B-2-2   & 0.3 ($k=1,..,5$), 0.25 ($k=6,..,10$) & RD=0.2 ($k=1, 2, 3$), =0.1 ($k=4, 5, 6$), =0($k=7,..,10$)         \\
        &               &            &                                                         &                                                    \\
{\it aRR}      & 6             & RR-A-1-1   & (0.15, 0.15, 0.1, 0.1, 0.05, 0.05)                      & (1.3, 1.3, 1.3, 1.3, 1.3, 1.3)                     \\
        &               & RR-A-1-2   & (0.35, 0.35, 0.3, 0.3, 0.2, 0.2)                        & (1.3, 1.3, 1.3, 1.3, 1.3, 1.3)                     \\
        &               & RR-A-2-1   & (0.15, 0.15, 0.1, 0.1, 0.05, 0.05)                      & (1.3, 1.3, 1.3, 1, 1, 1)                           \\
        &               & RR-A-2-2   & (0.35, 0.35, 0.3, 0.3, 0.2, 0.2)                        & (1.3, 1.3, 1.3, 1, 1, 1)                           \\
        &               &            &                                                         &                                                    \\
        & 10            & RR-B-1-1   & 0.1 (k=1,..,5), 0.05 (k=6,..,10) & RR=1.3 for all baskets \\
        &               & RR-B-1-2   & 0.3 (k=1,..,5), 0.25 (k=6,..,10) & RR=1.3 for all baskets \\
        &               & RR-B-2-1   &  0.1 (k=1,..,5), 0.05 (k=6,..,10)  & RR=1.3 ($k=1, 2, 3$), =1.1 ($k=4, 5, 6$), =1 ($k=7,..,10$)         \\
        &               & RR-B-2-2   & 0.3 (k=1,..,5), 0.25 (k=6,..,10) & RR=1.3 ($k=1, 2, 3$), =1.1 ($k=4, 5, 6$), =1 ($k=7,..,10$)  \\
\hline       
\end{tabular}
\end{center}
\end{table}
\end{landscape}

\begin{table}[]
\caption[]
	{\textit{Summary of the simulation study for $K=6$ to evaluate performance of the MH estimators; {\it CP} implies the coverage probability and on the row {\it size}, empirical coverage probabilities for the one-tailed 2.5 percent nominal level are presented for the asymptotic and the exact tests.  
	}}
\begin{center}
\small
\begin{tabular}{ccccccccc}
\hline
     &    &            &         &           &         &        & \multicolumn{2}{c}{Size} \\
     & $K$  & dataset    & true & estimator & average & {\it CP}(\%) & Asymptotic       & Exact       \\
\hline
{\it Null} & 6  & Null-A-1-1 & 0       & MH-RD     & 0.000   & 90.5   & 4.8        & 1.6         \\
     &    &            & 1       & MH-RR     & 1.002   & 90.5   & 4.8        & 1.6         \\
     &    &            &         & MH-iwRR   & 1.001   & 90.5   & 4.7        & 2.3         \\
     &    & Null-A-1-2 & 0       & MH-RD     & 0.000   & 92.9   & 3.5        & 2.3         \\
     &    &            & 1       & MH-RR     & 0.999   & 92.9   & 3.5        & 2.3         \\
     &    &            &         & MH-iwRR   & 0.999   & 92.7   & 3.7        & 2.5         \\
     & 10 & Null-B-1-1 & 0       & MH-RD     & -0.001  & 94.0   & 3.0        & 2.1         \\
     &    &            & 1       & MH-RR     & 0.996   & 94.0   & 3.0        & 2.1         \\
     &    &            &         & MH-iwRR   & 0.996   & 94.0   & 3.0        & 2.6         \\
     &    & Null-B-1-2 & 0       & MH-RD     & 0.000   & 95.2   & 2.4        & 2.1         \\
     &    &            & 1       & MH-RR     & 1.000   & 95.2   & 2.4        & 2.1         \\
     &    &            &         & MH-iwRR   & 0.999   & 94.7   & 2.7        & 2.5         \\
     &    &            &         &           &         &        &            &             \\
{\it aRD}   & 6  & RD-A-1-1   & 0.100   & MH-RD     & 0.101   & 93.9   &            &             \\
     &    & RD-A-1-2   & 0.100   & MH-RD     & 0.100   & 95.0   &            &             \\
     &    & RD-A-2-1   & 0.069   & MH-RD     & 0.069   & 93.8   &            &             \\
     &    & RD-A-2-2   & 0.069   & MH-RD     & 0.070   & 94.9   &            &             \\
     &    &            &         &           &         &        &            &             \\
     & 10 & RD-B-1-1   & 0.100   & MH-RD     & 0.100   & 94.8   &            &             \\
     &    & RD-B-1-2   & 0.100   & MH-RD     & 0.100   & 94.3   &            &             \\
     &    & RD-B-2-1   & 0.093   & MH-RD     & 0.093   & 94.7   &            &             \\
     &    & RD-B-2-2   & 0.093   & MH-RD     & 0.092   & 94.3   &            &             \\
     &    &            &         &           &         &        &            &             \\
{\it aRR}   & 6  & RD-A-1-1   & 1.300   & MH-RR     & 1.302   & 94.8   &            &             \\
     &    &            &         & MH-iwRR   & 1.305   & 91.8   &            &             \\
     &    & RD-A-1-2   & 1.300   & MH-RR     & 1.300   & 94.6   &            &             \\
     &    &            &         & MH-iwRR   & 1.301   & 95.0   &            &             \\
     &    & RD-A-2-1   & 1.104   & MH-RR     & 1.151   & 94.2   &            &             \\
     &    &            &         & MH-iwRR   & 1.108   & 90.5   &            &             \\
     &    & RD-A-2-2   & 1.104   & MH-RR     & 1.115   & 95.4   &            &             \\
     &    &            &         & MH-iwRR   & 1.100   & 94.7   &            &             \\
     &    &            &         &           &         &        &            &             \\
     & 10 & RD-B-1-1   & 1.300   & MH-RR     & 1.300   & 94.4   &            &             \\
     &    &            &         & MH-iwRR   & 1.297   & 93.3   &            &             \\
     &    & RD-B-1-2   & 1.300   & MH-RR     & 1.299   & 94.6   &            &             \\
     &    &            &         & MH-iwRR   & 1.299   & 94.8   &            &             \\
     &    & RD-B-2-1   & 1.115   & MH-RR     & 1.138   & 94.1   &            &             \\
     &    &            &         & MH-iwRR   & 1.115   & 92.8   &            &             \\
     &    & RD-B-2-2   & 1.115   & MH-RR     & 1.122   & 94.9   &            &             \\
     &    &            &         & MH-iwRR   & 1.116   & 94.8   &            &         \\
\hline   
\end{tabular}
\end{center}
\end{table}

\begin{landscape}
\begin{table}[]
\caption[]
	{\textit{Summary of the simulation study for evaluating performance of GIC-based basket identification.
	}}
\scriptsize
\begin{center}
\begin{tabular}{ccccccccccccccccc}
\hline
         &          &  & \multicolumn{4}{c}{two subclasses} &  & \multicolumn{4}{c}{all subclasses} &  & \multicolumn{4}{c}{Non-sparse}     \\
Scenario & item     &  & 1      & 2    & 3     & 4     &  & 1      & 2      & 3     & 4     &  & 1     & 2     & 3     & 4      \\ \cline{1-2} \cline{4-7} \cline{9-12} \cline{14-17} 
1GN   &  & &\multicolumn{14}{c}{($\pi_1$, $\pi_2$, $\pi_3$, $\pi_4$)=(0.1, 0.1, 0.1, 0.1)}        \\

         & Estimate &  & 0.101   & 0.101 & 0.098  & 0.097  &  & 0.100   & 0.100   & 0.100  & 0.099  &  & 0.100  & 0.100  & 0.101  & 0.100   \\
         & $100 \times$Bias     &  & 0.084   & 0.104 & -0.162 & -0.269 &  & -0.020  & 0.014   & 0.032  & -0.075 &  & -0.012 & -0.037 & 0.063  & -0.018  \\
         & $100 \times$MSE      &  & 0.420   & 0.431 & 0.719  & 0.741  &  & 0.443   & 0.456   & 0.895  & 0.914  &  & 0.480  & 0.490  & 0.533  & 0.538   \\
         & $\%$Reject    &  & 1.7     & 1.6   & 1.6    & 1.4    &  & 0.4     & 0.3     & 0.2    & 0.4    &  & 1.0    & 1.1    & 1.1    & 1.3     \\

2GA   & &  &\multicolumn{14}{c}{($\pi_1$, $\pi_2$, $\pi_3$, $\pi_4$)=(0.3, 0.3, 0.3, 0.3)}       \\
         & Estimate &  & 0.299   & 0.301 & 0.301  & 0.300  &  & 0.299   & 0.301   & 0.302  & 0.299  &  & 0.300  & 0.301  & 0.300  & 0.299   \\
         & $100 \times$Bias     &  & -0.099  & 0.093 & 0.118  & 0.008  &  & -0.116  & 0.090   & 0.237  & -0.071 &  & -0.053 & 0.133  & 0.009  & -0.057  \\
         & $100 \times$MSE      &  & 1.003   & 1.007 & 1.675  & 1.723  &  & 1.062   & 1.056   & 2.054  & 2.082  &  & 1.144  & 1.152  & 1.230  & 1.219   \\
         & $\%$Reject    &  & 72.1    & 72.7  & 62.9   & 62.8   &  & 52.3    & 49.4    & 26.5   & 24.0   &  & 56.0   & 56.9   & 55.2   & 55.4    \\

3    &  & &\multicolumn{14}{c}{($\pi_1$, $\pi_2$, $\pi_3$, $\pi_4$)=(0.1, 0.1, 0.3, 0.3)}       \\
         & Estimate &  & 0.110   & 0.109 & 0.282  & 0.280  &  & 0.101   & 0.100   & 0.300  & 0.298  &  & 0.104  & 0.103  & 0.293  & 0.292   \\
         & $100 \times$Bias     &  & 0.996   & 0.906 & -1.794 & -1.968 &  & 0.099   & 0.018   & -0.024 & -0.167 &  & 0.445  & 0.332  & -0.718 & -0.794  \\
         & $100 \times$MSE      &  & 0.518   & 0.511 & 2.006  & 2.000  &  & 0.456   & 0.457   & 2.093  & 2.080  &  & 0.541  & 0.529  & 1.238  & 1.243   \\
         & $\%$Reject    &  & 6.6     & 6.3   & 40.5   & 40.3   &  & 0.7     & 0.7     & 16.1   & 16.2   &  & 3.5    & 3.3    & 43.8   & 43.6    \\

4    &  & &\multicolumn{14}{c}{($\pi_1$, $\pi_2$, $\pi_3$, $\pi_4$)=(0.1, 0.1, 0.1, 0.5)}       \\
         & Estimate &  & 0.104   & 0.103 & 0.114  & 0.476  &  & 0.100   & 0.100   & 0.101  & 0.502  &  & 0.117  & 0.123  & 0.216  & 0.306   \\
         & $100 \times$Bias     &  & 0.354   & 0.265 & 1.420  & -2.409 &  & 0.031   & -0.048  & 0.091  & 0.191  &  & 1.736  & 2.311  & 11.547 & -19.392 \\
         & $100 \times$MSE      &  & 0.437   & 0.424 & 0.815  & 3.195  &  & 0.456   & 0.448   & 0.912  & 2.507  &  & 1.008  & 1.074  & 3.660  & 4.509   \\
         & $\%$Reject    &  & 5.6     & 5.2   & 7.1    & 68.8   &  & 0.5     & 0.4     & 0.3    & 62.9   &  & 13.3   & 14.8   & 38.8   & 65.3    \\

5   &  & &\multicolumn{14}{c}{($\pi_1$, $\pi_2$, $\pi_3$, $\pi_4$)=(0.1, 0.5, 0.5, 0.5)}      \\
         & Estimate &  & 0.111   & 0.503 & 0.485  & 0.486  &  & 0.100   & 0.499   & 0.499  & 0.502  &  & 0.103  & 0.503  & 0.493  & 0.493   \\
         & $100 \times$Bias     &  & 1.091   & 0.247 & -1.466 & -1.380 &  & -0.014  & -0.087  & -0.131 & 0.161  &  & 0.297  & 0.329  & -0.727 & -0.696  \\
         & $100 \times$MSE      &  & 0.628   & 1.041 & 1.729  & 1.702  &  & 0.466   & 1.255   & 2.481  & 2.502  &  & 0.491  & 1.223  & 1.767  & 1.750   \\
         & $\%$Reject    &  & 5.8     & 97.9  & 92.3   & 92.6   &  & 0.6     & 94.5    & 63.1   & 63.4   &  & 1.5    & 94.6   & 91.3   & 91.5    \\

6    &  & &\multicolumn{14}{c}{($\pi_1$, $\pi_2$, $\pi_3$, $\pi_4$)=(0.1, 0.3, 0.3, 0.5)}        \\
         & Estimate &  & 0.120   & 0.311 & 0.306  & 0.436  &  & 0.100   & 0.302   & 0.302  & 0.499  &  & 0.102  & 0.318  & 0.354  & 0.411   \\
         & $100 \times$Bias     &  & 2.014   & 1.124 & 0.623  & -6.452 &  & 0.011   & 0.177   & 0.216  & -0.144 &  & 0.189  & 1.785  & 5.386  & -8.886  \\
         & $100 \times$MSE      &  & 0.617   & 1.459 & 2.027  & 2.527  &  & 0.457   & 1.059   & 2.042  & 2.485  &  & 0.466  & 1.244  & 2.643  & 1.872   \\
         & $\%$Reject    &  & 9.8     & 71.8  & 67.1   & 92.4   &  & 1.4     & 44.4    & 18.4   & 64.2   &  & 2.0    & 54.7   & 68.9   & 88.4    \\

7   &  & &\multicolumn{14}{c}{($\pi_1$, $\pi_2$, $\pi_3$, $\pi_4$)=(0.1, 0.3, 0.5, 0.7)}       \\
         & Estimate &  & 0.149   & 0.305 & 0.569  & 0.626  &  & 0.100   & 0.301   & 0.599  & 0.701  &  & 0.101  & 0.303  & 0.646  & 0.649   \\
         & $100 \times$Bias     &  & 4.872   & 0.473 & 6.906  & -7.446 &  & 0.036   & 0.049   & 9.906  & 0.077  &  & 0.066  & 0.281  & 14.608 & -5.151  \\
         & $100 \times$MSE      &  & 0.969   & 2.601 & 3.476  & 2.902  &  & 0.461   & 1.024   & 3.376  & 2.108  &  & 0.467  & 1.097  & 3.430  & 1.416   \\
         & $\%$Reject    &  & 18.5    & 54.8  & 96.6   & 99.5   &  & 1.2     & 41.1    & 83.7   & 95.3   &  & 1.2    & 41.0   & 99.1   & 99.8   \\
\hline
\end{tabular}
\end{center}
\end{table}
\end{landscape}

\clearpage 

\appendix

\section*{Appendix A: Asymptotic property of the MH-RD}
\subsection*{A-1: Consistency and asymptotic normality of $\hat{\Delta}^{RD}$ under Asymptotic 1}
By simple algebra, under Asymptotic 1,
\begin{align*}
\hat{\Delta}^{RD} &=  \frac{\sum_{k=1}^K (Y_k-n_k \pi_{k,0})}{\sum_{k=1}^K n_k } =
\frac{\sum_{k=1}^K (\frac{n_k}{n}\frac{Y_k}{n_k}-\frac{n_k}{n} \pi_{k,0})}{\sum_{k=1}^K  \frac{n_k}{n} }
\nonumber \\
&\overset{p}{\longrightarrow} \frac{\sum_{k=1}^K q_k (\pi_k-\pi_{k,0})}{\sum_{k=1}^K  q_k } 
=\frac{\sum_{k=1}^K q_k \Delta_k^{RD}}{\sum_{k=1}^K q_k }
\end{align*}
The last quantity is denoted by $\Delta_*^{RD}$.
If the common risk difference assumption (\ref{rd1}) holds, it agrees with $\Delta^{RD}$. 

By simple algebra, from the central limit theorem, it holds that
\begin{align}
\sqrt{n} (\hat{\Delta}^{RD}-\Delta_*^{RD}) &= 
\frac{\sum_{k=1}^K \sqrt{\frac{n_k}{n}} \sqrt{n_k}(\hat{\pi_k}-\pi_k)}{\sum_{k=1}^K \frac{n_k}{n}}
\nonumber \\
&\overset{d}{\longrightarrow}  N \Big(0, \frac{\sum_{k=1}^K q_k2 \pi_k(1-\pi_k)}{(\sum_{k=1}^K q_k)^2}\Big),
\tag{A.1}
\label{nrd1}
\end{align}
where $\overset{d}{\longrightarrow}$ implies convergence in distribution.

\subsection*{A-2: Consistency and asymptotic normality of $\hat{\Delta}^{RD}$ under Asymptotic 2}
Under Asumptotic 2, there are finete number of configrations of $\{n_k\}$. Let the number of the configrations denoted by $L$. The $n_k$ for the $l$th configration is denoted by $\tilde{n}_l$ and the number of strata in the $l$th configration is denoted by $K_l$. It is supposed that $\lim_{K \to \infty} K_l/K$ exists and in $(0,1)$. Then, the denominator of the MH-RD estimator is represented as 
\begin{align*}
\sum_{l=1}^L \frac{K_l}{K} K_l^{-1} \sum_{j=1}^{K_l} \tilde{n}_l =
\sum_{l=1}^L \frac{K_l}{K} \tilde{n}_l.
\end{align*}
It converges to a non-zero constant as $K \to \infty$. \\
Under Asymptotic 2, $R_k^{RD}=Y_k-n_k \pi_{k,0}$ is uniformly bounded over $k=1,2,...$. Then, by the Kolmogorov's strong law of large number (Corollary of Theorem 5.4.1 of Chung (1974)), it holds that
\begin{align*}
& \lim_{K \to \infty} K^{-1} \sum_{k=1}^K 
\{(Y_k-n_k \pi_{k,0})-(E(Y_k)-n_k \pi_{k,0})\} \\
& = 
\lim_{K \to \infty} K^{-1} \sum_{k=1}^K 
\{(Y_k-n_k \pi_{k,0})-n_k \Delta_k^{RD})
\}=0
\end{align*}
almost surely. Then, if $\lim_{K \to \infty} K^{-1} \sum_{k=1}^K n_k \Delta_k^{RD}$ exists, it holds
\begin{align*}
\hat{\Delta}^{RD} &\overset{p}{\longrightarrow}  
\frac{\lim_{K \to \infty} K^{-1} \sum_{k=1}^K n_k \Delta_k^{RD}}
{\lim_{K \to \infty} K^{-1} \sum_{k=1}^K n_k }
\tag{A.2}
\label{av_rd2}
\end{align*}
Let the last quantity (\ref{av_rd2}) denoted by $\Delta_{**}^{RD}$.
If the common risk difference assumption (\ref{rd1}) holds, there exists $\lim_{K \to \infty} K^{-1} \sum_{k=1}^K n_k \Delta_k^{RD}$ and (\ref{av_rd2}) agrees with $\Delta^{RD}$. 

Since $\hat{\Delta}^{RD}$ is the solution to $U^{RD}(\Delta)=0$ and $\Delta_{**}^{RD}$ is the solution to the limit of $K^{-1} U^{RD}(\Delta)$, 
with simple algebraic manipulation and the central limit theorem, it holds that
\begin{align}
\sqrt{K}(\hat{\Delta}^{RD}-\Delta_{**}^{RD}) &= 
\Big(\frac{1}{K} \sum_{k=1}^K S_k^{RD}\Big)^{-1} \frac{1}{\sqrt{K}} U^{RD}(\Delta_{**}^{RD})+o_p(1)
\nonumber \\
&\overset{d}{\longrightarrow} N\Big(0, \frac{\sigma_{RD}^2}{(\lim_{k \to \infty} K^{-1} \sum_{k=1}^K S_k)^2}\Big),
\tag{A.3}
\label{nrd2}
\end{align} 
where 
\begin{align*}
\sigma_{RD}^2 &= \lim_{K \to \infty} K^{-1} \sum_{k=1}^K Var(R_k^{RD}-\Delta_{**}^{RD} S_k^{RD})
\nonumber \\
&= \lim_{K \to \infty} K^{-1} \sum_{k=1}^K n_k \pi_k (1-\pi_k).
\end{align*}

\subsection*{A-3: Derivation and dual consistency of $V\hat{a}r(\hat{\Delta}^{RD})$}
Simple algebraic manipulation entails that
\begin{align}
E\Big\{\frac{n_k}{n_k-1} \hat{\pi_k}(1-\hat{\pi_k})\Big\}=\pi_k (1-\pi_k).
\tag{A.4}
\label{unbiased}
\end{align}
Then, $V\hat{a}r(R_k^{RD}-\Delta_{**}^{RD}S_k^{RD})=n_k^2 \hat{\pi_k}(1-\hat{\pi_k})/(n_k-1)$ is an unbiased estimator for $Var(R_k^{RD}-\Delta_{**}^{RD}S_k^{RD})$. Replacing  $Var(R_k^{RD}-\Delta_{**}^{RD}S_k^{RD})$ in (\ref{nrd2}) with this unbiased estimator, the variance estimator (\ref{v2_rd}) is obtained.
Then, under Asymptotic 2, 
\begin{align*} 
\lim_{K \to \infty} K \times V\hat{a}r(\hat{\Delta}^{RD} ) &= 
\lim_{K \to \infty} \frac{K^{-1} \sum_{k=1}^K \frac{n_k^2}{n_k-1} E\{\hat{\pi}_k(1-\hat{\pi}_k)\}}
{(\lim_{K \to \infty} K^{-1} \sum_{k=1}^K S_k^{RD})^2} \nonumber \\
&= \lim_{K \to \infty} \frac{K^{-1} \sum_{k=1}^K \frac{n_k^2}{n_k-1} \frac{n_k-1}{n_k} \pi_k(1-\pi_k)\}}
{(\lim_{K \to \infty} K^{-1} \sum_{k=1}^K S_k^{RD})^2} 
\nonumber \\
&= \lim_{K \to \infty} \frac{K^{-1} \sum_{k=1}^K n_k \pi_k(1-\pi_k)\}}
{(\lim_{K \to \infty} K^{-1} \sum_{k=1}^K S_k^{RD})^2} 
\end{align*}
This agrees with the asymptotic variance (\ref{nrd2}). Then, $V\hat{a}r(\hat{\Delta}^{RD} )$ is consistent under Asymptotic 2. 

From (\ref{nrd1}), 
\begin{align}
\lim_{n \to \infty} n Var{(\hat{\Delta}^{RD})} &= 
\frac{\sum_{k=1}^K q_k \pi_k(1-\pi_k)}{(\sum_{k=1}^K q_k )^2}
\tag{A.5}
\label{vlimrd1}
\end{align}
On the other hand, 
\begin{align}
\lim_{n \to \infty} n V\hat{a}r(\hat{\Delta}^{RD} ) &= 
\lim_{n \to \infty} \frac{n^{-1}\sum_{k=1}^K \frac{n_k^2}{n_k-1} \hat{\pi}_k(1-\hat{\pi}_k)}
{(n^{-1}\sum_{k=1}^K S_k^{RD})^2} \nonumber \\
&=
\lim_{n \to \infty} \frac{\sum_{k=1}^K \frac{n_k}{n_k-1} \frac{n_k}{n} \hat{\pi}_k(1-\hat{\pi}_k)}
{(\sum_{k=1}^K \frac{n_k}{n})^2}  \nonumber \\
&=\frac{\sum_{k=1}^K q_k \pi_k(1-\pi_k)}
{(\sum_{k=1}^K q_k)^2}, 
\nonumber
\end{align}
where the last equivalence holds under Asymptotic 1. 
It agrees with (\ref{vlimrd1}), which implies dual consistency of the variance estimator (\ref{v2_rd}).

\section*{Appendix B: Asymptotic property of the MH-RR}
\subsection*{B-1: Consistency and asymptotic normality of $\hat{\Delta}^{RR}$ under Asymptotic 1}
Under Asymptotic 1, it holds that
\begin{align}
\hat{\Delta}^{RR} &= \frac{\sum_{k=1}^n w_k \frac{n_k}{n}\frac{Y_k}{n_k}}{\sum_{k=1}^n w_k \frac{n_k}{n} \pi_{k,0}} 
\nonumber \\
&\overset{p}{\longrightarrow} \frac{\sum_{k=1}^n w_k q_k \pi_k}{\sum_{k=1}^n w_k q_k \pi_{k,0}} =
\frac{\sum_{k=1}^n w_k q_k \pi_{k,0} \Delta_k^{RR}}{\sum_{k=1}^n w_k q_k \pi_{k,0}} 
\nonumber
\end{align}
The last quantity is denoted by $\Delta_*^{RR}$. If the common risk ratio assumption (\ref{rr1}) holds, it agrees with $\Delta^{RR}$. 

From the central limit theorem, it holds that
\begin{align}
\sqrt{n} (\hat{\Delta}^{RR}-\Delta_*^{RR}) &= 
\frac{\sum_{k=1}^K \sqrt{\frac{n_k}{n}} w_k \sqrt{n_k}(\hat{\pi}_k-\pi_k)}{\sum_{k=1}^K \frac{n_k}{n} w_k \pi_{k,0}}
\nonumber \\
&\overset{d}{\longrightarrow} N\Big(0, \frac{\sum_{k=1}^K q_k w_k^2 \pi_k(1-\pi_k)}{(\sum_{k=1}^K q_k w_k \pi_{k,0})^2}\Big)
\tag{B.1}
\label{nrr1}
\end{align}

\subsection*{B-2: Consistency and asymptotic normality of $\hat{\Delta}^{RR}$ under Asymptotic 2}
We begin with the case of $w_k=\pi_{k,0}^{-1}$. That is, we consider the MH-iwRR estimator;
\begin{align}
\hat{\Delta}^{RR}=\frac{K^{-1} \sum_{k=1}^K Y_k}{K^{-1}\sum_{k=1}^K n_k}. 
\tag{B.2}
\label{rr001}
\end{align}
As shown in Appendix A-3, the denominator (\ref{rr001}) converges to a non-zero constant as $K \to \infty$. 
By the Kolmogorov's strong law of large number (Corollary of Theorem 5.4.1 of Chung (1974)), it holds that
\begin{align*}
& \lim_{K \to \infty} K^{-1} \sum_{k=1}^K 
\{Y_k-E(Y_k)\} 
 = 
\lim_{K \to \infty} K^{-1} \sum_{k=1}^K 
\{Y_k-n_k \Delta_k^{RR})
\}=0
\end{align*}
almost surely. Then, if $\lim_{K \to \infty} K^{-1} \sum_{k=1}^K n_k \Delta_k^{RR}$ exists, it holds
\begin{align*}
\hat{\Delta}^{RR} &\overset{p}{\longrightarrow}  
\frac{\lim_{K \to \infty} K^{-1} \sum_{k=1}^K n_k \Delta_k^{RR}}
{\lim_{K \to \infty}  K^{-1} \sum_{k=1}^K n_k } 
\tag{B.3}
\label{av_rr2}
\end{align*}
If the common RR assumption (\ref{rr1}) holds, there exists $\lim_{K \to \infty} K^{-1} \sum_{k=1}^K n_k \Delta_k^{RR}$ and then
(\ref{av_rr2}) agrees with $\Delta^{RR}$. 

Next we consider the case of general weights other that $w_k=\pi_{k,0}^{-1}$. That is, we consider the following estimator;
\begin{align}
\hat{\Delta}^{RR}=\frac{K^{-1} \sum_{k=1}^K w_k Y_k}{K^{-1}\sum_{k=1}^K w_k n_k \pi_{k,0}}. 
\tag{B.4}
\label{rr002}
\end{align}
We assume an additional condition; there exists only finite number of weights $\{w_k\}$ and $\{\pi_{k,0}\}$ over $k=1,2,...$. Under this condition, there are only finite number of configurations for $w_k n_k \pi_{k,0}$. Then, in a similar argument to the case of $w_k=\pi_{k,0}^{-1}$, one can show that the denominator of (\ref{rr002}) converges to a non-zero constant. 
By the Kolmogorov's strong law of large number (Corollary of Theorem 5.4.1 of Chung (1974)), it holds that
\begin{align*}
& \lim_{K \to \infty} K^{-1} \sum_{k=1}^K 
\{w_k Y_k-w_k E(Y_k)\} 
 = 
\lim_{K \to \infty} K^{-1} \sum_{k=1}^K 
w_k \{Y_k-n_k \pi_{k,0} \Delta_k^{RR})
\}=0
\end{align*}
If $\lim_{K \to \infty} K^{-1} \sum_{k=1}^K n_k \pi_{k,0} \Delta_k^{RR}$ exists, it holds that
\begin{align*}
\hat{\Delta}^{RR} &\overset{p}{\longrightarrow}  
\frac{\lim_{K \to \infty}K^{-1} \sum_{k=1}^K n_k w_k \pi_{k,0} \Delta_k^{RR}}
{\lim_{K \to \infty}K^{-1}  \sum_{k=1}^K n_k w_k \pi_{k,0}} 
\tag{B.5}
\label{av_rr3}
\end{align*}
Let (\ref{av_rr2}) or (\ref{av_rr3}) denoted by $\Delta_{**}^{RR}$ in each setting.
If the common RR assumption (\ref{rr1}) holds, it agree with $\Delta^{RR}$. 

Since $\hat{\Delta}^{RR}$ is the solution to $U^{RR}(\Delta)=0$ and $\Delta_{**}^{RR}$ is its limit, 
with simple algebraic manipulation and the central limit theorem, it holds that
\begin{align}
\sqrt{K}(\hat{\Delta}^{RR}-\Delta_{**}^{RR}) &=
\Big(\frac{1}{K} \sum_{k=1}^K S_k^{RR}\Big)^{-1} \frac{1}{\sqrt{K}} U^{RD}(\Delta_{**}^{RR})+o_p(1) 
\nonumber \\
&\overset{d}{\longrightarrow}   N\Big(0, \frac{\sigma_{RR}^2}{(\lim_{k \to \infty} K^{-1} \sum_{k=1}^K S_k^{RR})^2}\Big),
\tag{B.6}
\label{nrr2}
\end{align} 
where 
\begin{align*}
\sigma_{RR}^2 &= \lim_{K \to \infty} K^{-1} \sum_{k=1}^K Var(R_k^{RR}-\Delta_{**}^{RR} S_k^{RR})
\nonumber \\
&= \lim_{K \to \infty} K^{-1} \sum_{k=1}^K w_k^2 \frac{n_k^2}{n_k-1} \pi_k (1-\pi_k).
\end{align*}

\subsection*{B-3: Derivation and dual consistency of $V\hat{a}r(\hat{\Delta}^{RR})$}
The variance estimator (\ref{v2_rr}) is constructed in a similar way to Appendix A-3. 
Then, 
\begin{align*} 
\lim_{K \to \infty} K \times V\hat{a}r(\hat{\Delta}^{RR} ) &= 
\lim_{K \to \infty} \frac{K^{-1} \sum_{k=1}^K \frac{n_k^2}{n_k-1} w_k^2 E\{\hat{\pi}_k(1-\hat{\pi}_k)}
{(\lim_{K \to \infty}K^{-1} \sum_{k=1}^K S_k^{RR})^2} \nonumber \\
&= \lim_{K \to \infty}\frac{ K^{-1} \sum_{k=1}^K \frac{n_k^2}{n_k-1} w_k^2 \frac{n_k-1}{n_k} \pi_k(1-\pi_k)}
{(\lim_{K \to \infty}  K^{-1} \sum_{k=1}^K S_k^{RR})^2} 
\nonumber \\
&= \lim_{K \to \infty} \frac{\sum_{k=1}^K n_k w_k^2 \pi_k(1-\pi_k)}
{(\lim_{K \to \infty} \sum_{k=1}^K S_k^{RR})^2} 
\end{align*}
This agrees with the asymptotic variance (\ref{nrr2}) under Asymptotic 2.

From (\ref{nrr1}), 
\begin{align}
\lim_{n \to \infty} n Var{(\hat{\Delta}^{RR})} &= \frac{\sum_{k=1}^K q_k w_k^2 \pi_k(1-\pi_k)}{(\sum_{k=1}^K q_k w_k \pi_{k,0})^2}.
\tag{B.7}
\label{vlimrr1}
\end{align}
On the other hand, 
\begin{align}
\lim_{n \to \infty} n V\hat{a}r(\hat{\Delta}^{RR} ) &= 
\lim_{n \to \infty} \frac{n^{-1}\sum_{k=1}^K \frac{n_k^2}{n_k-1} w_k^2 \hat{\pi}_k(1-\hat{\pi}_k)}
{(n^{-1}\sum_{k=1}^K S_k^{RR})^2} \nonumber \\
&=
\lim_{n \to \infty} \frac{\sum_{k=1}^K \frac{n_k}{n_k-1} \frac{n_k}{n}w_k^2 \hat{\pi}_k(1-\hat{\pi}_k)}
{(\sum_{k=1}^K w_k \frac{n_k}{n})^2}  \nonumber \\
&= \frac{\sum_{k=1}^K q_k w_k^2 \pi_k(1-\pi_k)}
{(\sum_{k=1}^K w_k q_k \pi_{k,0})^2}, 
\nonumber
\end{align}
where the last equivalence holds under Asymptotic 1. 
It agrees with $(\ref{vlimrr1})$, which implies dual consistency of the variance estimator (\ref{v2_rr}).

\section*{Appendix C: Assessing the common parameter assumptions}
In the standard Mantel-Haenszel procedure for the stratified $2\times 2$ tables, the common parameter assumption is often addressed with the Brewlow-Day test (Breslow and Day 1980). For the one-sample case, we can apply the standard chi-squared goodness-of-fit test. Suppose the common RD assumption. The basket-specific response rate is predicted as $\tilde{\pi}_k=\pi_{k,0}+\hat{\Delta}^{RD}$. Then, the chi-squared goodness-of fit statistics is given as
\begin{align*}
Z_{GOF}^2=\sum_{k=1}^K \frac{(Y_k-n_k \tilde{\pi}_k)^2}{n_k \tilde{\pi}_k}.
\end{align*}
Referring it with chi-squared distribution with the degree of freedom $K-1$, one can test the common RD assumption. A similar test can be applied for the common RR assumption.

\section*{Appendix D: Generalized information criterion}
\subsection*{D-1: Derivation of the GIC}
The bias term (\ref{bias}) of the GIC is derived from Theorem 3.1 of Konishi and Kitagawa (1996) for the general M-estimators applied to the estimating equation (\ref{eq1}). This is a justification of the GIC under Asymptotic 2. Hereafter, we show that the bias term (\ref{bias}) is also justified under Asymptotic 1.

Denote $\bm Y=(Y_1, Y_2, ..., Y_K)$ and $\bm y=(y_1, y_2, ..., y_K)$. The probability mass function for $Y_k$ is 
\begin{align*}
f_k(y_k)&=P(Y_k=y_k; \pi_{k,0}, \Delta)=\frac{n_k \!}{y_k \! (n_k-y_k)\!} \pi_k^{y_k} (1-\pi_k)^{n_k-y_k} \nonumber \\
&= \frac{n_k \!}{y_k \! (n_k-y_k)\!} h_k(\Delta)^{y_k} (1-h_k(\Delta))^{n_k-y_k}
\end{align*}
The joint probability mass function is given by $f(\bm y; \Delta)=\prod_{k=1}^K f_k(y_k)$. 
Then, the log-likelihood function for $\bm y$ is given by
\begin{align*}
l(\bm Y; \Delta) &= \sum_{k=1}^K \{Y_k \log h_k(\Delta)+(n_k-Y_k) \log (1-h_k(\Delta))\} \nonumber \\
&= \sum_{k=1}^K l_k(Y_k; \Delta). 
\end{align*}
Let $\tilde{\bm Y}=(\tilde{Y}_1, \tilde{Y}_2, ..., \tilde{Y}_K)$ be an independent sample from $f(\bm y; \Delta)=\prod_{k=1}^K f_k(y_k)$, which represents future observation. Following the idea of the information criterion, the goodness of a model is measured by 
\begin{align}
-\tilde{E}[E\{\log f(\bm \tilde{Y}; \hat{\Delta})\}]=-\tilde{E}[E\{l(\bm Y; \hat{\Delta})\}],
\tag{D.1}
\label{exploglik}
\end{align}
where $E$ implies the expectation with respect to $\bm Y$, which is data used for estimating $\hat{\Delta}$, and  $\tilde{E}$ is with respect to $\bm \tilde{Y}$. The model minimizing (\ref{exploglik}) is regarded as minimizing the Kullback-Leibler divergence over data (Konishi and Kitagawa 1996). 
If we estimate $\tilde{E}[E\{\log f(\bm \tilde{Y}; \hat{\Delta})\}]$ with $\log f(\bm Y; \hat{\Delta})$, it has bias since $\bm Y$ is dependent on $\hat{\Delta}$. A bias-adjusted estimate is an information criterion. We evaluate the bias hereafter. It is
\begin{align}
bias&= -\tilde{E}[E\{\log f(\tilde{\bm Y}; \hat{\Delta})\}]+E\{\log f(\bm Y; \hat{\Delta})\} \nonumber \\
&= -\sum_{k=1}^K \Big\{ \tilde{E}[E\{\log f(\tilde{Y}_k; \hat{\Delta})\}]-E\{\log f(Y_k; \hat{\Delta})\} \Big\} \nonumber \\
&= -\sum_{k=1}^K b_k. \nonumber
\end{align}
The bias $b_k$ is decomposed as
\begin{align}
b_k &=  E\Big[\tilde{E}\big\{\log f(\tilde{Y}_k; \hat{\Delta})\big\}-\log f(Y_k; \hat{\Delta})\Big]  \nonumber \\
&=  E\Big[\tilde{E}\big\{\log f(\tilde{Y}_k; \hat{\Delta})\big\}-\tilde{E}[\log f(\tilde{Y}_k; \Delta)\big\}\Big]
\tag{D.2}
  \label{t1} \\
&+  E\Big[\tilde{E}\big\{\log f(\tilde{Y}_k; \Delta)\big\}-\log f(Y_k; \Delta)\Big] 
\tag{D.3} \label{t2}  \\
&+  E\Big[\log f(Y_k; \Delta)-\log f(Y_k; \hat{\Delta})\Big] 
\tag{D.4}
\label{t3}
\end{align}
By simple algebra, one can see that (\ref{t2})=0.  
\begin{align*}
(\ref{t1}) &= E\Big[\tilde{E}\big\{\log{\frac{n_k !}{\tilde{Y}_k ! (n_k-\tilde{Y}_k)! }}+
\tilde{Y}_k \log{h_k(\hat{\Delta})}+(n_k-\tilde{Y}_k) \log{(1-h_k(\hat{\Delta}_k))}\big\}\Big] \nonumber \\
& - E\Big[\tilde{E}\big\{\log{\frac{n_k !}{\tilde{Y}_k ! (n_k-\tilde{Y}_k)!}}+
\tilde{Y}_k \log h_k(\Delta)+(n_k-\tilde{Y}_k) \log (1-h_k(\Delta))\big\}\Big] 
\nonumber \\
&= E \Big[n_k \pi_k \big\{\log{h_k(\hat{\Delta})}-\log{h_k(\Delta)}\big\}
+n_k(1-\pi_k)[\log{(1-h_k(\hat{\Delta}_k))}-\log (1-h_k(\Delta))\big\}\Big] \nonumber \\
&= E \Big[ n_k \pi_k \frac{\dot{h}_k(\Delta)}{h_k(\Delta)}
\big\{\hat{\Delta}-\Delta+O_p(n^{-1})\big\}
+n_k(1-\pi_k) \frac{\dot{h}_k(\Delta)}{1-h_k(\Delta)}
\big\{\hat{\Delta}-\Delta+O_p(n^{-1})\big\}   \nonumber \\
&= O(1). \nonumber 
\end{align*}
The last identify holds since 
\begin{align*}
\hat{\Delta}-\Delta=\frac{\sum_{k=1}^K (R_k-\Delta S_k)}{\sum_{k=1} S_k}
\end{align*}
and then $E(\hat{\Delta}-\Delta)=0$.
\begin{align}
(\ref{t3}) &= -E\Big[\log{\frac{n_k !}{Y_k ! (n_k-Y_k)! }}+
Y_k \log{h_k(\hat{\Delta})}+(n_k-Y_k) \log{(1-h_k(\hat{\Delta}_k))}\Big] \nonumber \\
& + E\Big[\log{\frac{n_k !}{y_k ! (n_k-y_k)!}}+
y_k \log h_k(\Delta)+(n_k-y_k) \log (1-h_k(\Delta))\}  \nonumber \\ 
&= -E\Big[y_k \frac{\dot{h}_k(\Delta)}{h_k(\Delta)}(\hat{\Delta}-\Delta+O_p(n^{-1}))
+(n_k-Y_k) \frac{\dot{h}_k(\Delta)}{1-h_k(\Delta)}(\hat{\Delta}-\Delta+O_p(n^{-1}))\} \nonumber \\
&= -E\Big[\big\{Y_k \frac{\dot{h}_k(\Delta)}{h_k(\Delta)}
+(n_k-Y_k) \frac{\dot{h}_k(\Delta)}{1-h_k(\Delta)}\big\}
(\hat{\Delta}-\Delta) \Big]+O(1) \nonumber \\
&= -(\sum_{k=1}^K S_k)^{-1} E\Big[\big\{Y_k \frac{\dot{h}_k(\Delta)}{h_k(\Delta)}
+(n_k-Y_k) \frac{\dot{h}_k(\Delta)}{1-h_k(\Delta)} \big\} 
\sum_{j=1}^K (R_j-\Delta S_j) \Big]+O(1) \nonumber \\
&= -(\sum_{k=1}^K S_k)^{-1} E\Big[\big\{y_k \frac{\dot{h}_k(\Delta)}{h_k(\Delta)}
+(n_k-y_k) \frac{\dot{h}_k(\Delta)}{1-h_k(\Delta)}\big\} (R_k-\Delta S_k) \Big]+O(1).
\tag{D.5}
\label{bias11}
\end{align}
Note that under Asymptotic 1, the first term of the GIC (\ref{gic}) diverges. Ignoring $O(1)$ term in (\ref{bias11}) and summing (\ref{bias11}) up over $k=1,2,..,K$, one can obtain the bias term (\ref{bias}) of the GIC.

\subsection*{D-2: An intuitive explanation for effectivenss of GIC}
With a simple case, we explain that the bias term of the GIC actually protect unnecessary separation of baskets. We compare two models.
In Model A, we estimate the common RD (or RR) over all the baskets. In Model B, baskets are classified into two subclasses; we assume that the first subclass consists of the first $K_1$ baskets ($k=1, 2, \dots, K_1$) and the second one consists of the rest ($k=K_1+1, K_1+2\dots K$). The MH estimate with Model A is denoted by $\hat{\Delta}_A$. Then, the GIC for Model A is written by
\begin{align*}
GIC_A=-\sum_{k=1}^K \Big[ Y_k log h_k(\hat{\Delta}_A+(n_k-Y_k)log(1-h_k(\hat{\Delta}_A))\Big]+bias(\hat{\Delta}_A),
\end{align*}
where 
\begin{align}
bias(\hat{\Delta}_A) = (\sum_{k=1}^K S_k)^{-1} \sum_{k=1}^K 
(R_k-\hat{\Delta}_A S_k)
\Big\{
Y_k \frac{\dot{h}_k(\hat{\Delta}_A)}{h_k(\hat{\Delta}_A)}-(n_k-Y_k)\frac{\dot{h}_k(\hat{\Delta}_A)}{1-h_k(\hat{\Delta}_A)}
\Big\}
\tag{D.6}
\label{bias_a}
\end{align}
Let the MH estimates for the first and second subclasses of Model B denoted by $\hat{\Delta}_{B,1}$ and $\hat{\Delta}_{B,2}$, respectively. The GIC for Model B is given by
\begin{align}
GIC_B=-\sum_{k=1}^{K_1} \Big[ Y_k log h_k(\hat{\Delta}_{B,1}+(n_k-Y_k)log(1-h_k(\hat{\Delta}_{B,1}))\Big]+bias_{B,1}(\hat{\Delta}_{B,1}) \nonumber \\
-\sum_{k=K_1+1}^{K} \Big[ Y_k log h_k(\hat{\Delta}_{B,2}+(n_k-Y_k)log(1-h_k(\hat{\Delta}_{B,2}))\Big]+bias_{B,2}(\hat{\Delta}_{B,2}),
\nonumber 
\end{align}
where 
\begin{align}
bias_{B,1}(\hat{\Delta}_{B,1}) &= (\sum_{k=1}^{K_1} S_k)^{-1} \sum_{k=1}^{K_1} (R_k-\hat{\Delta}_{B,1} S_k)
\Big\{
Y_k \frac{\dot{h}_k(\hat{\Delta}_{B,1})}{h_k(\hat{\Delta}_{B,1})}-(n_k-Y_k)\frac{\dot{h}_k(\hat{\Delta}_{B,1})}{1-h_k(\hat{\Delta}_{B,1})}
\Big\} \nonumber \\
bias_{B,2}(\hat{\Delta}_{B,2}) &= (\sum_{k=K_1+1}^K S_k)^{-1} \sum_{k=K_1+1}^K (R_k-\hat{\Delta}_{B,2} S_k)
\Big\{
Y_k \frac{\dot{h}_k(\hat{\Delta}_{B,2})}{h_k(\hat{\Delta}_{B,2})}-(n_k-Y_k)\frac{\dot{h}_k(\hat{\Delta}_{B,2})}{1-h_k(\hat{\Delta}_{B,2})}
\Big\}
\nonumber
\end{align}
Suppose Model A is correct. In words, the common RD assumption holds. Under this assumption, all of $\hat{\Delta}_{A}$, $\hat{\Delta}_{B,1}$ and $\hat{\Delta}_{B,2}$ estimate the common RD consistently. Ideally, we suppose that $\hat{\Delta}_{A}=\hat{\Delta}_{B,1}=\hat{\Delta}_{B,2}$. Then, the GIC for Molde B becomes
\begin{align*}
GIC_B&=-\sum_{k=1}^{K} \Big[ Y_k log h_k(\hat{\Delta}_{A}+(n_k-Y_k)log(1-h_k(\hat{\Delta}_{A}))\Big] \nonumber \\
&+bias_{B,1}(\hat{\Delta}_{A})+bias_{B,2}(\hat{\Delta}_{A}).
\end{align*}
The likelihood part agrees with that of Model 1. The bias term of Model B is given by
\begin{align}
 & bias_{B,1}(\hat{\Delta}_{A})+bias_{B,2}(\hat{\Delta}_{A}) \nonumber \\
&=
(\sum_{k=1}^{K_1} S_k)^{-1} \sum_{k=1}^{K_1} (R_k-\hat{\Delta}_{A} S_k)
\Big\{
Y_k \frac{\dot{h}_k(\hat{\Delta}_{A})}{h_k(\hat{\Delta}_{A})}-(n_k-Y_k)\frac{\dot{h}_k(\hat{\Delta}_{A})}{1-h_k(\hat{\Delta}_{A})}
\Big\} 
\tag{D.7} \label{bias_b1} \\
& + (\sum_{k=K_1+1}^K S_k)^{-1} \sum_{k=K_1+1}^K (R_k-\hat{\Delta}_{A} S_k)
\Big\{
Y_k \frac{\dot{h}_k(\hat{\Delta}_{A})}{h_k(\hat{\Delta}_{A})}-(n_k-Y_k)\frac{\dot{h}_k(\hat{\Delta}_{A})}{1-h_k(\hat{\Delta}_{A})}
\Big\}.
\tag{D.8}
\label{bias_b2}
\end{align}
The bias term (\ref{bias_a}) of Model A is represented as
\begin{align}
& bias_{A}(\hat{\Delta}_{A}) \nonumber \\
&=
(\sum_{k=1}^{K} S_k)^{-1} \sum_{k=1}^{K_1} (R_k-\hat{\Delta}_{A} S_k)
\Big\{
Y_k \frac{\dot{h}_k(\hat{\Delta}_{A})}{h_k(\hat{\Delta}_{A})}-(n_k-Y_k)\frac{\dot{h}_k(\hat{\Delta}_{A})}{1-h_k(\hat{\Delta}_{A})}
\Big\} 
\tag{D.9}
\label{bias_a1} \\
&+ (\sum_{k=1}^K S_k)^{-1} \sum_{k=K_1+1}^K (R_k-\hat{\Delta}_{A} S_k)
\Big\{
Y_k \frac{\dot{h}_k(\hat{\Delta}_{A})}{h_k(\hat{\Delta}_{A})}-(n_k-Y_k)\frac{\dot{h}_k(\hat{\Delta}_{A})}{1-h_k(\hat{\Delta}_{A})}
\Big\}.
\tag{D.10}
\label{bias_a2}
\end{align}
The inequalities $(\ref{bias_a1}) \le (\ref{bias_b1})$ and  $(\ref{bias_a2}) \le (\ref{bias_b2})$ hold and then $GIC_A \le GIC_B$. Thus, the GIC protects unnecessary separation of baskets.

\section*{Appendix E: One-sample Mantel-Haenszel odds ratio}
Denote the odds ratio of the $k$th basket by
\begin{align*}
\Delta_k^{OR}=\frac{(1-\pi_{k,0})\pi_k}{\pi_{k,0}(1-\pi_k)}.
\end{align*}
The common odds ratio assumption is given by $\Delta_1^{OR}=\Delta_2^{OR}=...=\Delta_K^{OR}$ and the common value is denoted by $\Delta^{OR}$. Under the common odds ratio assumption, $\Delta^{OR} \pi_{k,0} (1-\pi_k)=(1-\pi_{k,0}) \pi_k$ holds for $k=1,2,...,K$. This relationship motivates us to use the one-sample Mantel-Haenszel odds ratio estimator
\begin{align*}
\hat{\Delta}^{OR} = \frac{\sum_{k=1}^K w_k (1-\pi_{k,0})y_k}{\sum_{k=1}^K w_k \pi_{k,0}(n_k-y_k)}.
\end{align*}
It is regarded as the solution to the estimating equation $U^{OR}(\Delta^{OR})=0$, where
\begin{align*}
U^{OR}(\Delta^{OR}) &= \sum_{k=1}^K 
w_k\{(1-\pi_{k,0})y_k - \Delta^{OR} \pi_{k,0} (n_k-y_k)\} \nonumber \\
&= 
\sum_{k=1}^K w_k(1-\pi_{k,0})y_k 
-\Delta^{OR} \sum_{k=1}^K w_k \pi_{k,0} (n_k-y_k) \nonumber \\
&=
\sum_{k=1}^K R_k^{OR}-\Delta^{OR} \sum_{k=1}^K S_k^{OR}, \nonumber
\end{align*}
where $R_k^{OR}=w_k(1-\pi_{k,0})y_k$ and $S_k^{OR}=w_k \pi_{k,0} (n_k-y_k)$. 
Under Asymptotic 2,  
$\sqrt{K}(\hat{\Delta}^{OR}-\Delta^{OR}) \simeq \{K^{-1}dU^{OR}(\Delta^{OR})/d\Delta^{OR}\}^{-1} K^{-\frac{1}{2}}U^{OR}(\Delta^{OR})$ holds and then $\sqrt{K}(\hat{\Delta}^{OR}-\Delta^{OR})$ asymptotically follows a zero-mean normal distribution with the variance
\begin{align*}
\frac{\lim_{K \to \infty} K^{-1} \sum_{k=1}^K Var(R_k^{OR}-\Delta^{OR}S_k^{OR}) }
{\{\lim_{K \to \infty} K^{-1} \sum_{k=1}^K S_k^{OR}\}^2} &=
\frac{\lim_{K \to \infty} K^{-1} \sum_{k=1}^K \{1+(\Delta^{OR}-1) \pi_{k,0}\}^2 n_k \pi_k (1-\pi_k) }
{\{\lim_{K \to \infty} K^{-1} \sum_{k=1}^K S_k^{OR}\}^2}.
\end{align*}
By (\ref{unbiased}), 
\begin{align}
V\hat{a}r(\hat{\Delta}^{OR}) = \frac{\sum_{k=1}^K \{1+(\hat{\Delta}^{OR}-1) \pi_{k,0}\}^2 \frac{n_k^2}{n_k-1} \hat{\pi}_k (1-\hat{\pi}_k) }{\{\sum_{k=1}^K S_k^{OR}\}^2}
\tag{E.1}
\label{v_or}
\end{align}
is motivated. In a similar way to the MH-RD and MH-RR, the dual consistency of (\ref{v_or}) can be shown.

The GIC is given by (\ref{gic}) with $h_k(x)=\pi_{k,0} x/(1-\pi_{k,0}+\pi_{k,0}x)$, $\hat{\Delta}=\hat{\Delta}^{OR}$, $R_k=R_k^{OR}$ and $S_k=S_k^{OR}$.

\end{document}